\documentclass[iop]{emulateapj}
\usepackage{hyperref}
\usepackage{xcolor}
\usepackage{url}
\usepackage{amsmath}

\newcommand{\hcn}{HCN($J = 2 \to 1$)}
\newcommand{\hnc}{HNC($J = 2 \to 1$)}
\newcommand{\hcop}{HCO$^{+}$($J = 2 \to 1$)}
\newcommand{\hto}{H$_{2}$O($3_{13} \to 2_{20}$)}
\newcommand{\hctn}{HC$_{3}$N($J = 20 \to 19$)}
\newcommand{\hiz}{high-$z$ }
\newcommand{\lco}{$L'_{\mathrm{CO}}$}
\newcommand{\lhcn}{$L'_{\mathrm{HCN}}$}
\newcommand{\lhnc}{$L'_{\mathrm{HNC}}$}
\newcommand{\lhcop}{$L'_{\mathrm{HCO^{+}}}$}
\newcommand{\lfir}{$L_{\mathrm{FIR}}$}
\newcommand{\mum}{$\mu$m}

\slugcomment{Accepted to the ApJ}

\shorttitle{}
\shortauthors{}

\begin{document}

\title{High dense gas fraction in a gas-rich star-forming galaxy at \lowercase{$z = 1.2$} }

\def\andname{\hspace*{-0.5em}}

\author{Avani Gowardhan \altaffilmark{1,*}, Dominik A. Riechers \altaffilmark{1}, Emanuele Daddi \altaffilmark{2},  Riccardo Pavesi \altaffilmark{1}, Helmut Dannerbauer \altaffilmark{3,4} and Chris Carilli \altaffilmark{5}}

\altaffiltext{*}{Email: ag2255@cornell.edu}
\altaffiltext{1}{Department of Astronomy, Cornell University, Ithaca, NY 14853, USA}
\altaffiltext{2}{CEA Saclay, Laboratoire AIM-CNRS-Universit\'e Paris Diderot, Irfu/SAp, Orme des Merisiers, 91191 Gif-sur-Yvette, France}
\altaffiltext{3}{Instituto de Astrof\'isica de Canarias (IAC), E-38205 La Laguna, Tenerife, Spain}
\altaffiltext{4}{Universidad de La Laguna, Dpto. Astrof\'isica, E-38206 La Laguna, Tenerife, Spain}
\altaffiltext{5}{National Radio Astronomy Observatory, PO Box O, Socorro, NM 87801, USA}
 
\begin{abstract}

We report observations of dense molecular gas in the star-forming galaxy EGS 13004291 ($z=1.197$) using the Plateau de Bure Interferometer. We tentatively detect HCN and HNC $J=2\to1$ emission when stacked together at $4\sigma$ significance, yielding line luminosities of $L'_{\mathrm{HCN}(J=2\to1)}=(9\pm3)\times10^{9}$ K km s$^{-1}$pc$^{2}$ and $L'_{\mathrm{HNC}(J=2\to1)}=(5\pm2)\times10^{9}$ K km s$^{-1}$pc$^{2}$ respectively. We also set 3$\sigma$ upper limits of $<7$--$8\times10^{9}$ K km s$^{-1}$pc$^{2}$ on the \hcop, H$_{2}$O($3_{13}\to2_{20}$) and HC$_{3}$N($J=20\to19$) line luminosities. We serendipitously detect CO emission from two sources at $z\sim1.8$ and $z\sim3.2$ in the same field of view. We also detect CO($J=2\to1$) emission in EGS 13004291, showing that the excitation in the previously detected CO($J=3\to2$) line is subthermal ($r_{32}=0.65\pm0.15$). We find a line luminosity ratio of \lhcn/\lco $=0.17\pm0.07$, as an indicator of the dense gas fraction. This is consistent with the median ratio observed in $z>1$ galaxies (\lhcn/\lco $=0.16\pm0.07$) and nearby ULIRGs (\lhcn/\lco$=0.13\pm0.03$), but higher than in local spirals (\lhcn/\lco$=0.04\pm0.02$). Although EGS 13004291 lies significantly above the galaxy main sequence at $z\sim1$, we do not find an elevated star formation efficiency (traced by \lfir/\lco) as in local starbursts, but a value consistent with main-sequence galaxies. The enhanced dense gas fraction, the subthermal gas excitation, and the lower than expected star formation efficiency of the dense molecular gas in EGS 13004291 suggest that different star formation properties may prevail in \hiz starbursts. Thus, using \lfir/\lco$ $ as a simple recipe to measure the star formation efficiency may be insufficient to describe the underlying mechanisms in dense star-forming environments inside the large gas reservoirs.

\end{abstract}

\keywords{galaxies -- evolution, galaxies -- high-redshift, galaxies -- star formation, galaxies -- starburst, ISM -- molecules } 

\section{Introduction}\label{sec:intro}

\let\thefootnote\relax\footnotetext{Based on observations carried out under project ID U030 with the IRAM NOEMA Interferometer. IRAM is supported by INSU/CNRS (France), MPG (Germany) and IGN (Spain).} 

Over the past decade, there has been a surge in the number of galaxies at the peak epoch of star formation ($z \sim 1-3$) in which molecular gas, the fuel for star formation, has been detected \citep[see][for a review]{carilli2013}. This includes a sizable sample of ``normal'' star-forming galaxies (SFGs ;  \citealt{daddi2008, daddi2010a, tacconi2010, tacconi2013}); most of these fall on a tight relation between stellar mass ($M_{*}$) and star formation rate (SFR $\sim M_{*}^{p}, p = 0.6 - 0.9$), the so-called star-forming ``main-sequence'' (MS; \citealt{noeske2007, elbaz2007,daddi2007,wuyts2011a, rodighiero2011,whitaker2012}). The specific star formation rate (sSFR $=$ SFR$/M_{*}$) remains roughly constant along the MS at each epoch but increases with redshift, consistent with the overall increase in the cosmic star formation rate density \citep[e.g.,][]{karim2011, whitaker2012,speagle2014,lee2015}. The existence of the galaxy MS shows that the bulk of cosmic star formation at \hiz proceeds in a quasi-steady state, over long timescales ($\sim 1$ Gyr) and large spatial scales, while episodes of intense merger-induced starburst activity play a smaller role. Consequently, two main modes of star formation have been proposed for the \hiz SFG population : a quiescent star formation mode which dominates in MS galaxies, and an enhanced ``starburst'' mode of merger-driven star formation commonly seen in outliers lying above the MS \citep{daddi2010b,genzel2010, rodighiero2011}.  

MS galaxies at $z > 1$ have star formation rates that are $10-100$ times higher than SFRs observed in normal galaxies in the local Universe. A significant fraction of extreme \hiz galaxies such as sub-millimeter galaxies (SMGs) or merging ULIRGs are inconsistent with the MS \citep[see][for reviews]{blain2002, casey2014}, and display enhanced star formation efficiencies (SFE$_{\rm mol}$ =  SFR/$M_{\rm gas}$) and short gas depletion timescales ($\tau_{\rm depl}$ = 1/SFE$_{\rm mol}$). However, it is yet unclear what fraction of the enhanced star formation rate in \hiz SFGs is driven by their high gas masses, as opposed to an increase in their SFE, and whether this fraction differs between SFGs lying or above the galaxy MS  \citep{daddi2015, scoville2015, silverman2015}. While recent studies for above-MS galaxies at $z\sim 1.6$ have found a higher \lfir/\lco$ $ i.e. a higher SFE relative to MS galaxies, in 7 outliers above the MS \citep{silverman2015}, existing work is far from conclusive. Alternative theories suggest a single unified mode of intense star formation in all \hiz SFGs, driven by compression and turbulence in the interstellar medium (ISM), with constant replenishment of cold gas from the cosmic web \citep{bouche2010, lilly2013}. In this framework, all \hiz SFGs, including both MS and above-MS galaxies, have a near-constant gas depletion timescale \citep[e.g.,][]{scoville2015}. Such a mode would be fundamentally different from that seen in IR-luminous starburst galaxies, which boast concentrated and short-lived star formation activity, or that in local spiral galaxies, where star formation primarily takes place in clumps of giant molecular clouds (GMCs), distributed throughout the diffuse molecular gas. As the molecular gas is strongly coupled to the star formation, we need to better characterize its properties in galaxies lying on and above the MS to identify the prevalent mode of star formation in \hiz SFGs. 

The Kennicutt-Schmidt (KS) relation \citep{schmidt1959, kennicutt1998} relates the local average gas density to the local SFR, typically traced using the FIR luminosity \lfir. It also explains the observed correlation \lfir $\propto$ \lco$^{1.5}$ \citep[e.g.,][]{sanders1996,kewley2002}, where CO is typically used to trace molecular gas. However, ground-state CO transitions ($n_{\rm crit} \sim 10^{3}$ $\mathrm {cm}^{-3}$) trace both the dense and diffuse molecular gas due to their low critical density, while only the cold, dense component is immediately available for star formation. 

Most of the star formation in the Milky Way and the local Universe takes place in the dense cores of GMCs ($n_{\rm H_{2}} \sim 10^{5}$ cm$^{-3}$), which are better traced by dense gas tracers such as HCN, HCO$^{+}$ and HNC ($n_{\rm crit} \sim 10^{5}$ cm$^{-3}$) than CO. There exists a strong linear correlation between \lfir$ $ and \lhcn$ $ \citep{gao2004b}. This \lfir-\lhcn$ $ relation holds over 7 orders of magnitude, ranging from GMCs in the Milky Way to \hiz galaxies \citep{gao2004b,wu2005, gao2007, wu2010}. This is a much tighter relation than the \lfir-\lco$ $ relation, which also shows a change in slope depending on the sample selection \citep{sanders1996,solomon1997,gao2004b}. While such a relationship is shown by both HCN and HCO$^{+}$  \citep{riechers2006a,papadopoulos2007, gracia2006,garcia2012}, HCO$^{+}$ is more sensitive to ionization conditions and typically shows weaker emission than HCN. On the other hand, the HCN abundance can be enhanced by the presence of an Active Galactic Nucleus (AGN) or mechanically driven shocks or outflows, contributing to non-linearity in the \lhcn-\lfir$ $ relation for more FIR luminous sources \citep{garcia2012,privon2015,martin2015,izumi2016}. Together, the relative intensities of dense gas tracers, HCN, HNC and \hcop$ $ allow us to probe the gas temperature and ionization \citep{costagliola2011}. In addition, the line luminosity ratio \lhcn/\lco$ $ is an indicator of the dense gas fraction $f_{\rm dense}$, and represents the \emph{actively star-forming} fraction of molecular gas. As \lhcn/\lco, \lfir/\lhcn, and \lfir/\lco$ $ vary significantly between different star formation environments, these are critical tools for identifying the dominant mode of star formation in \hiz SFGs, and they provide significantly more powerful constraints on the global star formation mechanisms than \lfir/\lco$ $ alone.

However, studies of the dense molecular gas as traced by HCN require high observing sensitivity, and thus have been mostly limited to the nearby Universe. There are only 3 solid and 3 tentative detections at high-$z$, including lensed systems, (U)LIRGs and SMGs \citep[e.g.,][and references therein]{gao2007, riechers2007}; there are no such observations for normal, unlensed SFGs at high-$z$. 

In this paper, we present for the first time observations of the dense, actively star-forming gas in a massive star-forming galaxy, EGS 13004291, at high redshift ($z = 1.197$). EGS 13004291 is the most CO-luminous source in the currently known sample of \hiz SFGs \citep{daddi2010b,tacconi2013}, and is therefore the best candidate for this initial study of the dense gas fraction in \hiz SFGs. 

The paper is organized as follows: in \autoref{sec:obs} and \autoref{sec:results}, we present the observations and results. In \autoref{sec:analysis}, we discuss our analysis, including line stacking and spectral energy distribution (SED) fitting, as well as the serendipitous detection of two sources. In \autoref{sec:discussion} and \autoref{sec:conclusions}, we discuss our results and conclusions. We use a $\Lambda$CDM cosmology, with $H_{0} = 71$ km s$^{-1}$ Mpc$^{-1}$, $\Omega_{\rm M} = 0.27$, and $\Omega_{\Lambda} = 0.73$ \citep{spergel2007}. 

\section{Observations}\label{sec:obs}

\begin{table*}[]
\begin{center}
\caption{\textsc{Observed line properties for EGS 13004291}}
\label{tab:table1}
		\begin{tabular}{l|c|c|c|c|c}
			\hline
			\hline
			Transition 				& $\nu_{\rm rest}$ & $\nu_{\rm obs}$ & $I$ & $L'$ & Reference$^{a}$  \\
									& 	(GHz)		& (GHz)		& (Jy km s$^{-1}$)	&  ($10^{10}$ K km s$^{-1}$ pc$^{2}$) & \\
			\hline
			CO($ J = 2 \to 1$) 		& 230.5379  	& 104.961 	& $3.09 \pm 0.27$ 		& $6.0 \pm 0.5$ & [1] \\
			CO($ J = 3 \to 2$) 		& 345.7959  	& 157.394 	& $4.6 \pm 0.1$ 		& $3.9$	 		& [2]\\
			HCN($ J = 2 \to 1$) 	& 177.2612 	& 80.704 	& $0.28 \pm 0.09$ 		& $0.9 \pm 0.3$ & [1] \\
			HNC($ J = 2 \to 1$) 	& 181.3248 	& 82.555 	& $0.17 \pm 0.07$ 		& $0.5 \pm 0.2$ & [1]\\
			\hctn$^{b}$				& 181.9449 	& 82.837 	& $ < 0.21$ 			& $ < 0.65$ 	& [1] \\
			\hcop$^{b}$ 			& 178.3750 	& 81.212	& $ < 0.23$ 			& $ < 0.76$ 	& [1] \\
			\hto$^{b}$  				& 183.3101 	& 83.459 	& $ < 0.22$ 			& $ < 0.66$ 	& [1] \\
			\hline \noalign {\smallskip}
		\end{tabular}
		
	\textbf{Notes:} $^{a}$ [1] This work. [2] \cite{tacconi2013}. $^{b}$ $3\sigma$ upper limits obtained as described in \autoref{sec:dgobs}.  
	\end{center}

\end{table*}

\subsection{IRAM observations}\label{sec:2.1}

We observed the primary target EGS 13004291 (J2000 RA: 14h19m15.0s, Dec:  +52d49m30s, $z = 1.197$) with the IRAM Plateau de Bure Interferometer (PdBI) with 5 antennas in the compact D-configuration during five tracks in May-June 2010, with a total on-source time of 11.5 hours. Weather conditions were average or better for 3 mm observations, with a precipitable water vapour (PWV) of 6--7 mm for all tracks. The absolute flux scale was calibrated either on 3C273, MWC349, or 3C345. The source J1418+546 was used as a phase and bandpass calibrator. The WideX correlator (bandwidth $\sim 3.6$ GHz) was used to observe multiple molecular lines simultaneously. Observations were carried out in dual polarization mode, using a tuning frequency of 81.95 GHz and a binned spectral resolution of 3.9 MHz ($\sim 14.2$ km s$^{-1}$ at 81.95 GHz), covering the frequency range $\nu_{\rm obs} = 80.6833-83.4366$ GHz. This covered the redshifted \hcn, \hnc, \hcop, \hctn, and \hto $ $ lines at $z = 1.197$ (see \autoref{tab:table1} for rest frequencies). 

In a second tuning, we observed the CO($ J = 2 \to 1$) line ($\nu_{\rm rest} = 230.538$ GHz) in May 2010 for 0.6 hours on-source, under excellent weather conditions for 3\,mm observing, using 6 antennas and recording data from both the narrow-band (bandwidth $\sim$ 1.0 GHz) and the WideX correlator. We used a tuning frequency of 104.933 GHz and binned spectral resolutions of 10 and 2 MHz, corresponding to $28$ km s$^{-1}$ and $5.6$ km s$^{-1}$ at $105$ GHz respectively. The data were taken in dual polarization mode, using MWC349 as the absolute flux calibrator.

All observations were calibrated using the IRAM PdBI data reduction pipeline in {\tt CLIC} (Continuum and Line Interferometer Calibration)$^1$\footnote[1]{$^1$\url{http://www.iram.fr/IRAMFR/GILDAS}}, with subsequent additional flagging by hand. The absolute flux scale was calibrated to better than $20\%$ for both CO and HCN observations. 

The reduced visibility cube for the dense gas tracers was imaged using {\tt UV\_MAP}, with natural weighting and a pixel size of $0.5'' \times 0.5''$, and cleaned using the task {\tt CLEAN} with the Hogbom algorithm, after binning over a large velocity width, 40 channels at our spectral resolution (corresponding to $\sim 590$ km s$^{-1}$). The resulting cleaned image has an rms noise of 0.15 mJy beam$^{-1}$, and a synthesized beam size of $6.0'' \times 5.0''$, with a position angle (PA) of -72.3\degr. The antenna half-power beam width is $61''$, and primary beam correction was applied using the task {\tt PRIMARY}. 

The task {\tt UV\_MAP} was used to image the calibrated visibility CO($J = 2 \to 1$) cube, using natural weighting and a pixel size of $0.5'' \times 0.5''$; the clean map was made using the task {\tt CLEAN}, using the Hogbom algorithm in {\tt MAPPING} with natural weighting. For the CO observations, the cleaned image has an rms noise of 2.3 mJy beam$^{-1}$ in each channel (width $\sim 17$ km s$^{-1}$) and a synthesized beam size of $4.7''\times 3.8''$, with a position angle (PA) of -66.5\degr. The antenna half-power beam width is $49''$.

\subsection{Archival Data}\label{sec:arch}

EGS 13004291 is located in the Extended Groth Strip, and as such has rich multi-wavelength coverage from the All-Wavelength Extended Groth Strip International Survey (AEGIS). This includes observations with \emph{Chandra} (2-10 keV; \citealt{laird2009,nandra2015}), {\em Galaxy Evolution Explorer }(GALEX), {\em Hubble Space Telescope} (\emph{HST}) as part of the 3D-HST survey \citep{brammer2012,skelton2014}, \emph{Spitzer} IRAC \citep{barmby2008}, and \emph{Herschel} PACS/SPIRE observations, as part of the NEWFIRM Medium-band Survey (NMBS ; \citealt{davis2007,whitaker2011}). 
	
To enable a comparison of our source properties against different galaxy populations, we have compiled a large sample of sources from the literature, both local and at high-$z$, with extant observations in the FIR, CO \citep{weiss2003,greve2005,riechers2006b, weiss2007, riechers2009,danielson2011,riechers2011,thomson2012} and HCN \citep{solomon2003,vanden2004,isaak2004,gao2004a,wagg2005, carilli2005, evans2006, greve2006, gao2007,gracia2008, riechers2010, krips2010,garcia2012}. All obtained luminosities have been adjusted to the cosmology used here; HCN detections with single dish telescopes, pointed at galactic nuclear regions have been treated as lower limits on the HCN luminosity where appropriate.  We exclude sources with upper or lower limits on luminosities from all relevant calculations and sample averages, and only use sources which have solid detections. 

\begin{table*}[]
	\begin{center}
		\caption{\textsc{Observational Summary for CO lines}}
		\label{tab:table2}
		\begin{tabular}{l|c|c|c|c|c|c}
			
			\hline
			\hline
			Source      			& $z$							& Transition 			&  $\Delta v_{\rm FWHM}$ 	& $I_{\rm CO}$ 		& \lco 									& Reference	\\
								&								&						& 	(km s$^{-1}$)		 	& (Jy km s$^{-1}$) 		& (10$^{10}$ K km s$^{-1}$ pc$^{2}$) 	& 				\\
			\hline
			EGS 13004291     	& $1.1964 \pm 0.0001$		& $ J = 2 \to  1$		&  $340 \pm 32$            	&  $3.09 \pm 0.27$  	& $6.0 \pm 0.5$						& [1]   			\\
								& $1.197$ 	 					& $ J = 3 \to 2$ 		&  $311$ 			             &  $4.6 \pm 0.1$           & $3.9 \pm 0.1$     						& [2]			\\
			J141917+524921 	& $1.8016 \pm 0.0002 $  		& $ J = 2 \to 1 $ 		&  $481 \pm 112$			&  $0.75 \pm 0.11$      	& $ 3.1 \pm 0.5 $     					& [1]			\\
			J141912+524924  	& $3.2206 \pm 0.0002$   		& $ J = 3 \to 2 $		&  $233 \pm 54 $            	&  $0.59 \pm 0.09$ 		& $ 3.0 \pm 0.5 $    						& [1]			\\
			\hline  \noalign {\smallskip}
		\end{tabular}
		
		\textbf{Notes:} Details of observed CO transitions. [1] This work. [2] \cite{tacconi2013}.
	\end{center}

\end{table*}

\section{Results}\label{sec:results}

\subsection{CO($J = 2 \to 1$)}\label{sec:gobs}

\begin{figure*}
		\vspace{-5mm}
\begin{center}
	\includegraphics[width=0.415\textwidth]{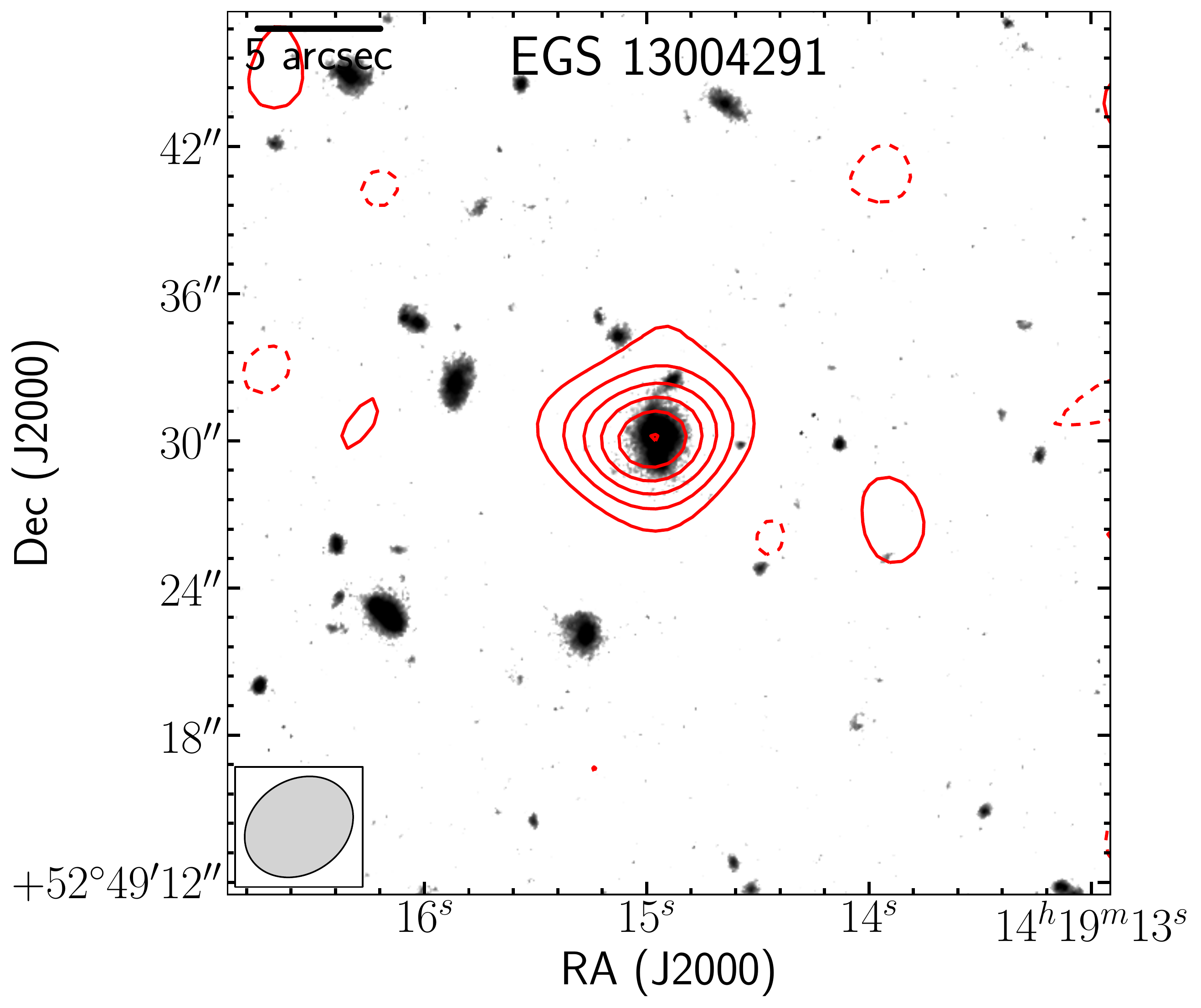}
	\includegraphics[width=0.575\textwidth]{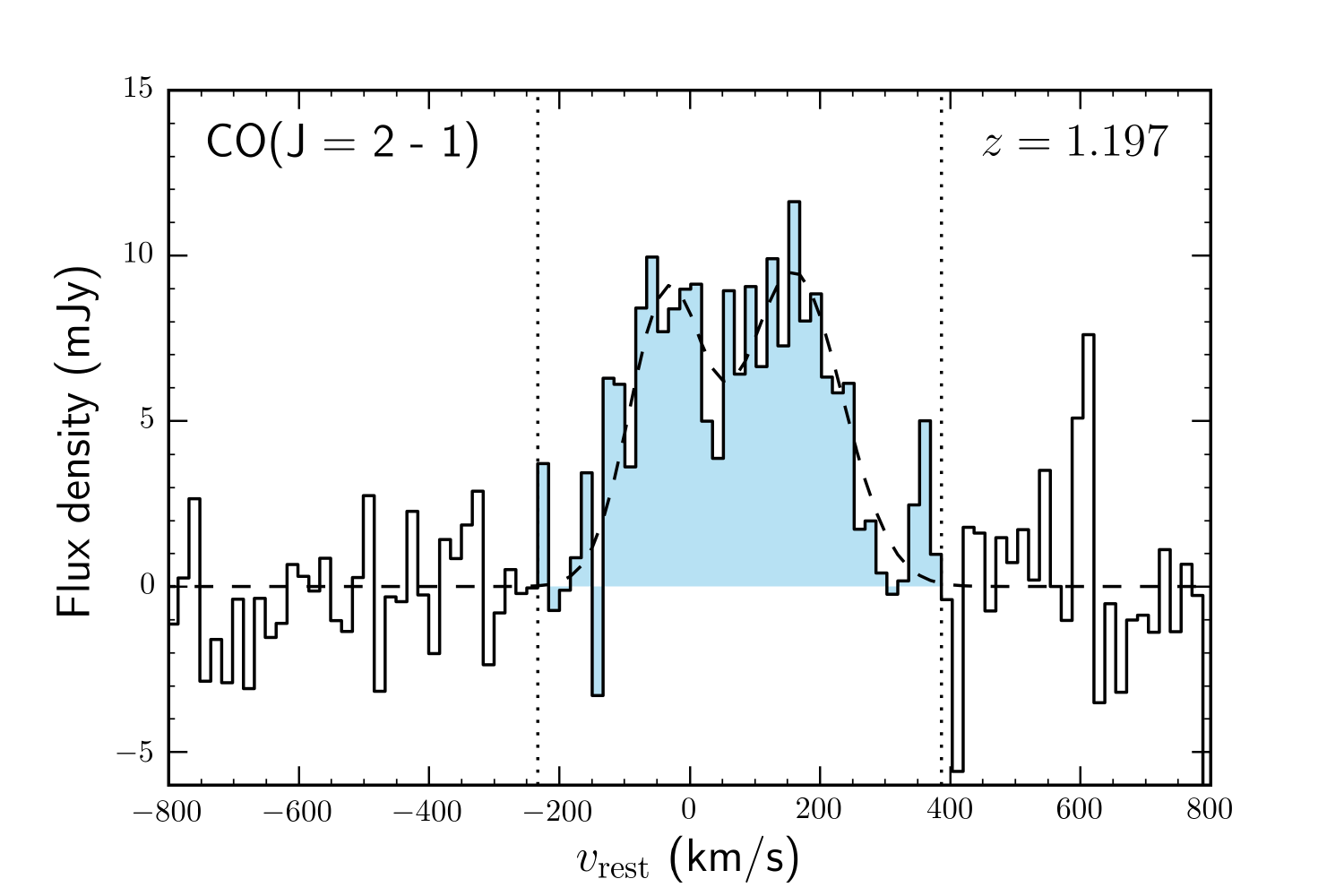}
	\caption{{\em Left}: CO($J = 2 \to 1$) moment-0 map for EGS 13004291, made by collapsing the cleaned cube along the frequency axis over the velocity range $-230$ km s$^{-1}$ to 386 km s$^{-1}$ (shown as dotted lines in spectrum). The source is detected at $\sim 12\sigma$ significance, where $1\sigma \sim 0.3$ Jy km s$^{-1}$. The relative contours at $\pm 2,4,6,8,10,12\sigma$ are shown, overlaid on the 3D-HST $H$ band image \citep{brammer2012,skelton2014}. The synthesized beam has a size of $\sim 4.7'' \times 3.8''$, and is indicated in the bottom left corner. {\em Right}: Spectrum extracted from the cleaned cube at the central pixel, with a velocity resolution of $\sim 17$ km s$^{-1}$.  The dashed line shows a 2-component Gaussian fit to the data. }
	\label{fig:co21}
	\end{center}
\end{figure*}

We successfully detect the CO($J = 2 \to 1$) line in EGS 13004291 at a redshift of $z = 1.197$. We find that the emission is spatially unresolved, consistent with what is expected based on the previously measured CO($J = 3 \to 2$) size (\citealt{tacconi2013};  $r_{1/2} \sim 0.5''$). We therefore extract the spectral profile (\autoref{fig:co21}) from the peak pixel (J2000 RA:  14h19m14.97s ; Dec: +52d49m29.73s). The spectrum is fit with a 1-D Gaussian to estimate the line peak, velocity width and central redshift; we find the line centre at $\nu_{\rm obs} = 104.961 \pm  0.005$ GHz corresponding to a redshift of $z = 1.1964 \pm 0.0001$. The moment-0 map (\autoref{fig:co21}) is created by collapsing the spectral cube along the frequency axis for the FWZI velocity width derived from this best fit ($\Delta v_{\rm FWZI} \sim 620$ km s$^{-1}$). Finally, the area under the spectral line fit is used to obtain an integrated line flux of $I_{\rm CO} = 3.09 \pm 0.27$ Jy km s$^{-1}$; this is consistent with the value derived from the peak of the moment-0 map. The final spectral profile is shown in \autoref{fig:co21}. From this, we find a CO($J = 2\to 1$) luminosity of $L'_{\mathrm{CO}(J = 2 \to 1)} =  (6.0 \pm 0.5) \times 10^{10}$ K km s$^{-1}$ pc$^{2}$. Assuming a brightness temperature ratio of $r_{21} = 0.76$ between the $J =2 \to 1$ and $J  = 1 \to 0$ transitions \citep{daddi2015}, we find a CO($J = 1 \to 0$) luminosity of $L'_{\mathrm{CO}(J = 1 \to 0)} =  (7.8 \pm 0.7) \times 10^{10}$ K km s$^{-1}$ pc$^{2}$. 

\begin{figure*}
	\begin{center}		
	\includegraphics[width=\textwidth]{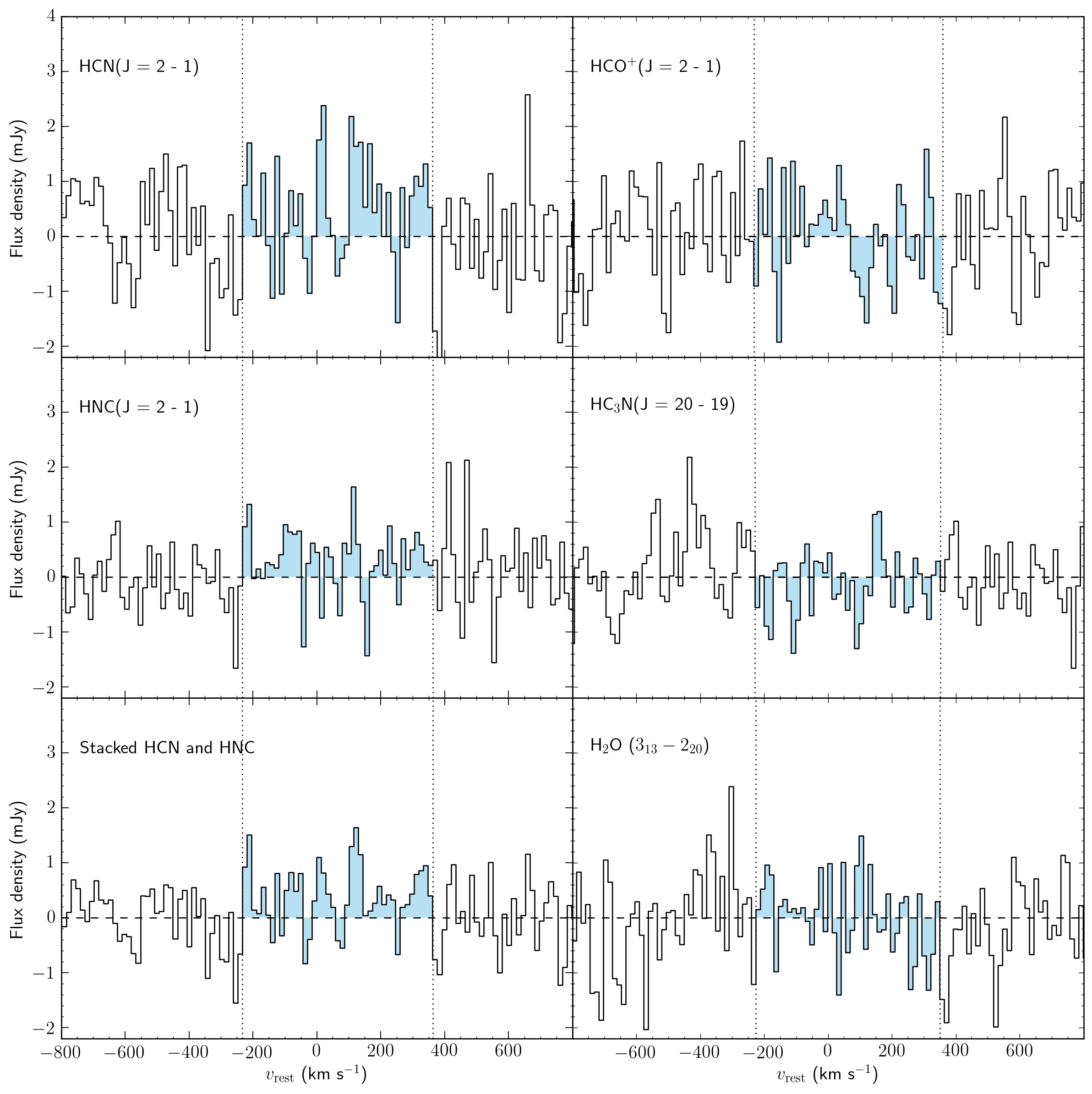}
	\caption{Observed spectra for HCN(\emph{top left}), HNC (\emph{middle left}), stacked HCN and HNC emission (\emph{bottom left}), HCO$^{+}$ (\emph{top right}), HC$_{3}$N (\emph{middle right}) and H$_{2}$O (\emph{bottom right}).  We tentatively detect HCN and HNC emission at $\sim 3\sigma$ and $\sim2.4\sigma$ significances respectively. By stacking the HCN and HNC lines together, we detect emission at $\sim 4\sigma$ significance. The other dense gas tracers including HCO$^{+}$, HC$_{3}$N and H$_{2}$O remain undetected. The dotted vertical lines show the regions used for calculating the velocity integrated line fluxes (or upper limits). }
	\label{fig:hcn_hnc_spec}
	\end{center}	
\end{figure*}

\begin{figure*}
	\begin{center}		
	\includegraphics[width=0.415\textwidth]{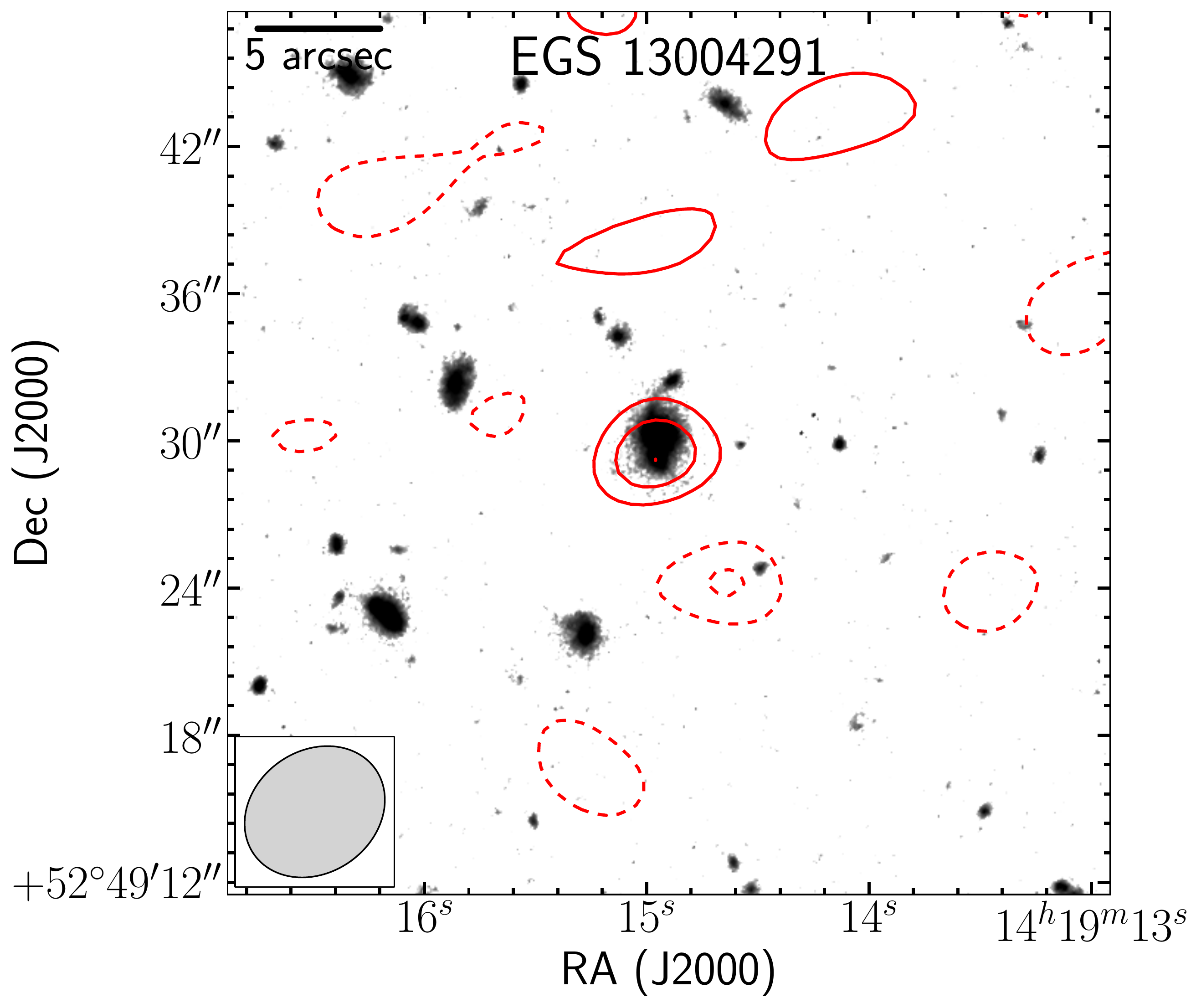}
	\includegraphics[width=0.575\textwidth]{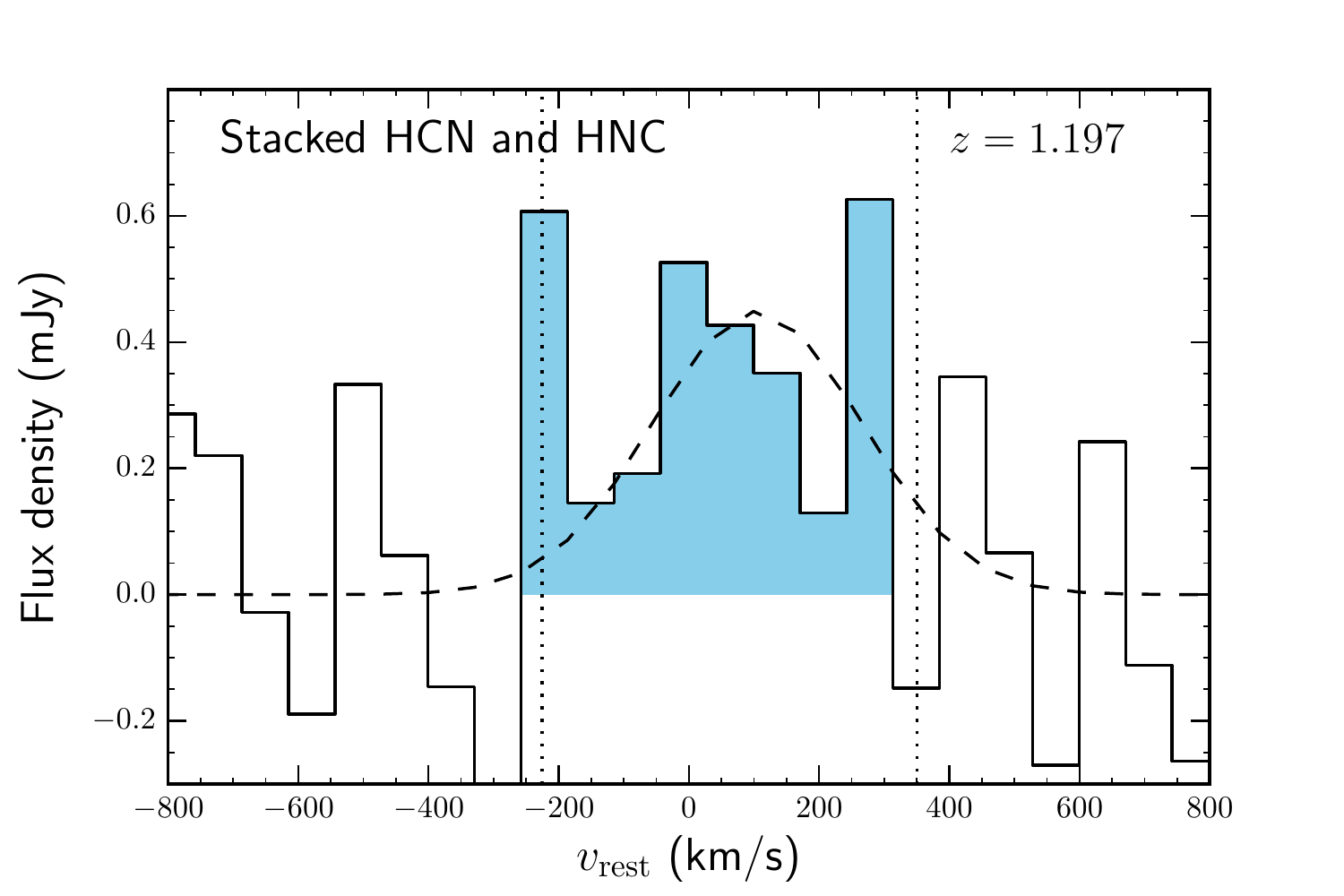}
	\caption{Stacked HCN and HNC emission, made by collapsing the reduced and cleaned stacked cube along the frequency axis over the velocity range $- 230$ to $360$ km s$^{-1}$ (shown as dotted lines in \autoref{fig:hcn_hnc_spec}) as for the CO($2 \to 1$) line in EGS 13004291. The stacked emission is detected at a $4\sigma$ significance, where $1\sigma \sim 0.05$ Jy km s$^{-1}$. The $\pm 2,3,4\sigma$ contours are overlaid on the 3D-HST $H$ band image \citep{brammer2012,skelton2014}. The synthesized beam has a size of $\sim 6.0'' \times 5.0''$, and is indicated in the bottom left corner. {\em Right:} Binned spectrum extracted from the stacked HCN and HNC cube, at the central pixel, with a velocity resolution of $\sim 72$ km s$^{-1}$. The dashed line shows a single-component Gaussian fit to the data.}
	\label{fig:hcn}
	\end{center}	
\end{figure*}

\subsection{HCN and HNC ($ J = 2 \to 1$)} \label{sec:dgobs}

For each of the lines, a moment-0 map is made using the velocity range $-230$ km s$^{-1}$ to $360$ km s$^{-1}$ ($\Delta v \sim 590$ km s$^{-1}$), nearly identical to the FWZI velocity range for the CO($J = 2 \to 1$) line, and the rms noise for each moment-0 map is calculated using the source-free pixels. We tentatively detect the HCN($ J = 2 \to 1$) and HNC($ J = 2 \to 1$) lines (\autoref{fig:hcn}), redshifted to $\nu_{\rm obs} = 80.704 $ GHz and $\nu_{\rm obs} = 82.555$ GHz respectively. Given the modest significance of our detections, we use the dirty cube for calculating integrated line fluxes, instead of the cleaned image cube in order to avoid biases introduced due to the cleaning process. We do not make any correction for flux in the sidelobes of the dirty beam, given that the source is spatially unresolved. We perform a 2-D Gaussian fit to the moment-0 maps of HCN and HNC emission in order to measure the line fluxes. We detect line emission at the source position at significances of $3.0\sigma$ and $2.4\sigma$ for HCN and HNC, with velocity integrated line fluxes of $I_{\rm HCN} = 0.28 \pm 0.09$ Jy km s$^{-1}$ and $I_{\rm HNC} = 0.17 \pm 0.07$ Jy km s$^{-1}$, respectively. These correspond to $L'_{\mathrm{HCN}(J = 2 \to 1)}$ $\sim (9 \pm 3) \times 10^{9}$ K km s$^{-1}$ pc$^{2}$ and $L'_{\mathrm{HNC}(J = 2 \to 1)}$ $ \sim (5 \pm 2) \times 10^{9}$ K km s$^{-1}$ pc$^{2}$. The HCN($J = 2 \to 1$) luminosity is extrapolated to the HCN($J = 1 \to 0$) luminosity using a brightness temperature ratio of $r_{21} \sim 0.65$ \citep{krips2008,geach2012} between the $J =2 \to 1$ and $J  = 1 \to 0$ transitions (see \autoref{sec:ahcn} for a discussion of how this impacts our results), which yields $L'_{\mathrm{HCN}(J = 1 \to 0)}$ $\sim (1.4 \pm 0.5) \times 10^{10}$ K km s$^{-1}$ pc$^{2}$. 

We use the same methodology as above to obtain upper limits on \hcop, \hctn, and \hto$ $ emission. We find $I \lesssim 0.2$ Jy km s$^{-1}$ pc$^{2}$ for each of the lines, yielding line luminosities $L' < 7$--$8 \times 10^{9}$ K km s$^{-1}$ pc$^{2}$ (\autoref{tab:table1}). These limits are consistent with the limits derived from stacking, as described in \autoref{sec:stacking}. No continuum emission was detected in the CO or dense gas observations, and we set a $3\sigma$ upper limit of $ < 0.3$ mJy beam$^{-1}$ on the continuum flux of the source at an observed frame wavelength of 3 mm. 

\section{Analysis}\label{sec:analysis}

\subsection{Stacking of dense gas tracers}\label{sec:stacking}

Since both the HCN and HNC lines are only tentatively detected at $ \lesssim 3\sigma$ significance, we stack these lines together using both image and $uv$-plane stacking to further investigate the reliability of their detection. We stack 200 velocity channels ($\sim$ 2900 km s$^{-1}$) around the predicted line peaks (using $\nu_{\rm obs}$ from \autoref{tab:table1} and $z = 1.197$) for the HCN and HNC($ J = 2 \to 1$) lines. For stacking in the visibility plane, the visibilities $V(u,v)$ are concatenated after the $u,v$ values are scaling by $\nu_{i}/\nu_{0}$, where $\nu_{i}$ is the frequency of channel $i$, and $\nu_{0}$ is the central frequency of observation. Dirty maps from the stacked visibilities are made using {\tt UV\_MAP}. For image plane stacking, the dirty spectral cubes for individual lines are stacked over the same channel width, centered on the expected HCN and HNC line centers. 

The spectral cubes, following both $uv$-plane and image-plane stacking, are binned over the velocity range $\Delta v \sim 590$ km s$^{-1}$ around the central channel, assuming that the velocity widths of CO, HCN, and HNC are the same. Consistency between the two methods is validated by comparing the rms noise achieved post-stacking. In addition, the weighted average of HCN and HNC line fluxes, as estimated in Section \autoref{sec:dgobs}, is consistent with the line flux detected in the stacked HCN and HNC emission map. The resulting map (\autoref{fig:hcn}) has an rms noise of $\sim 0.1$ mJy beam$^{-1}$, and an unresolved source is detected at the CO and optical position of EGS 13004291, with a peak flux of $0.4$ mJy, and at a $4\sigma$ significance.  

To ensure the validity of the approach, the same stacking routine is performed using central channels that have been randomly selected, rather than centered on known line positions. Dirty maps are made from the stacked visibilities, and searched for significant features. The rms noise is calculated per channel. The number of features found is consistent with distribution of the peak of a set of generated Gaussian images with an rms noise of $0.1$ mJy beam$^{-1}$ in each channel, as checked using a 2-sided Kolmogorov-Smirnov test. In a further 1000 trials, no features at an equal or higher significance are found within a synthesized beam size located at the central pixel. 

Finally, we also attempted stacking other dense gas tracers with HCN, and find that the significance of the detection is reduced in the stacked images. Removing the HCN contribution to the stacked image, we obtain $3\sigma$ upper limits on the HCO$^{+}$, HC$_{3}$N and H$_{2}$O line fluxes, consistent with those determined in Section \autoref{sec:dgobs}. We emphasize that for all our dense gas observations, we use stacking to confirm our results; however, our line fluxes and upper limits are determined for each dense gas tracer individually as described in Section \autoref{sec:dgobs}.

\subsection{Blind line search by Matched Filtering}\label{sec:mf}

Given the large bandwidth of the WideX correlator, we perform blind Matched Filtering on the reduced HCN image cube to search for other spectral lines in the primary beam. We use the code developed by Pavesi et al. (in prep.), which performs a blind search for spectral line features in interferometric data cubes. For a given source spatial extent and spectral line width, we generate a template spectral cube, assuming a 2-D circular Gaussian to describe the source structure and a 1-D Gaussian to describe the spectral line. Due to the spatial correlation of the noise, inherent in interferometric images, the source size is not trivially identical to the ``optimal" template size, but rather requires smaller templates instead (see Pavesi et al. for details).  We then convolve the templates defined in this way with the observed data cube, and the resulting cube is then searched for significant features. 

We perform the search for a range of possible template spatial sizes ($5'' < \Delta x < 17''$) and velocity widths ($100$ km s$^{-1}$ $ \lesssim \Delta v \lesssim  500$ km s$^{-1}$ ). While the same spectral feature might be detected at high significance in multiple templates with varying spatial extents, we expect the highest SNR when the source extent (after convolving with the telescope beam) ``matches" the template size, and therefore select the most significant detection of the source amongst all the templates. Finally, we obtain the spatial coordinates, peak channel, and significance of the features with the highest SNR. We visually check this list of features, removing obvious contaminants such as bad channels, and those features which occur at the edge of the image or of the spectral band. Finally, we check the remaining putative features for counterparts in the 3D-HST WFC3-selected photometric catalog of the AEGIS field and the multiwavelength AEGIS catalog \citep{davis2007, whitaker2011, brammer2012,skelton2014}. Here we discuss the two most likely candidates, both of which are detected at $\geq 6\sigma$ significance. 

\subsubsection{EGS J141917.4+524922}

\begin{figure*}
	\begin{center}
	\includegraphics[width=0.40\textwidth]{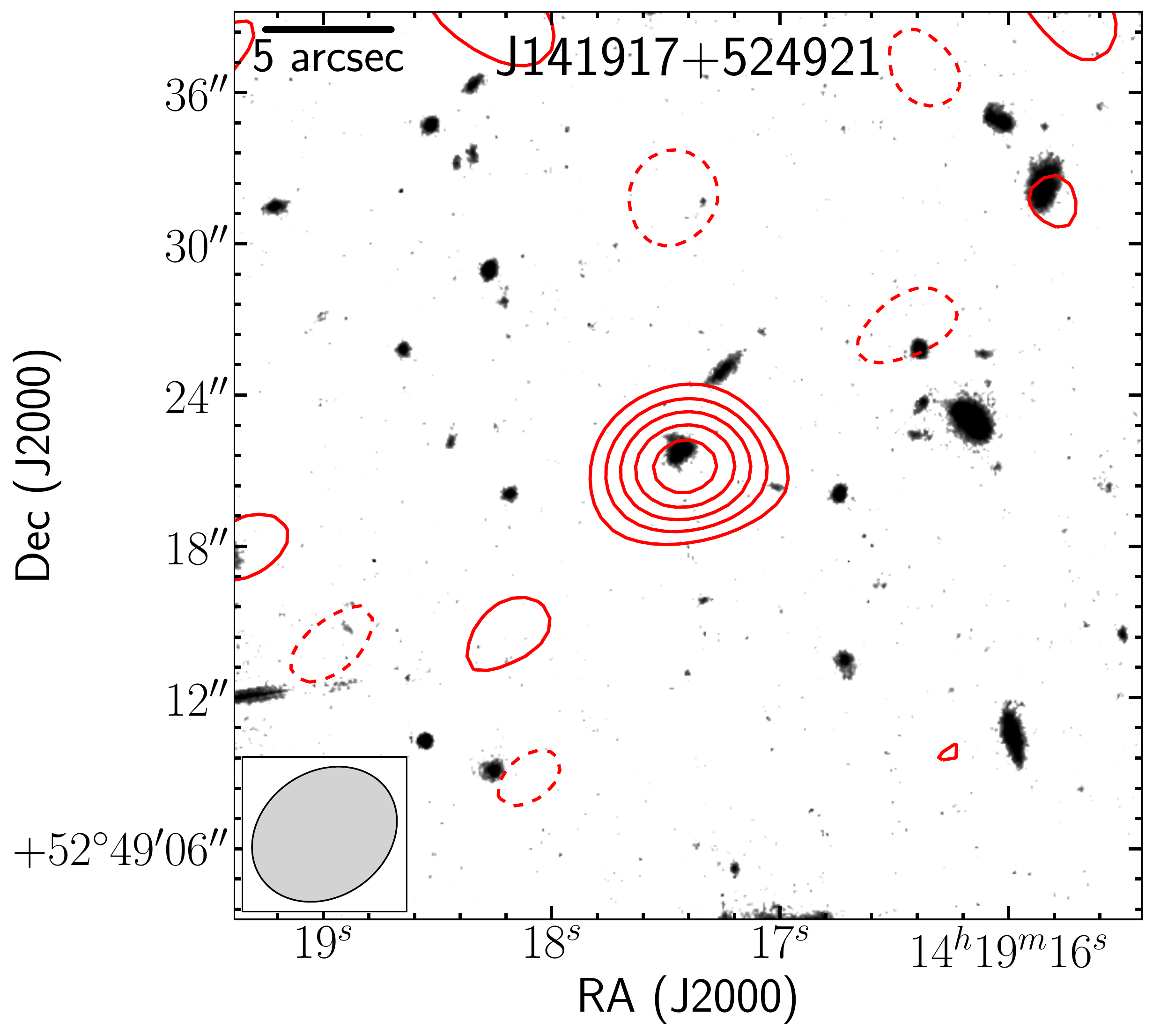}
	\includegraphics[width=0.58\textwidth]{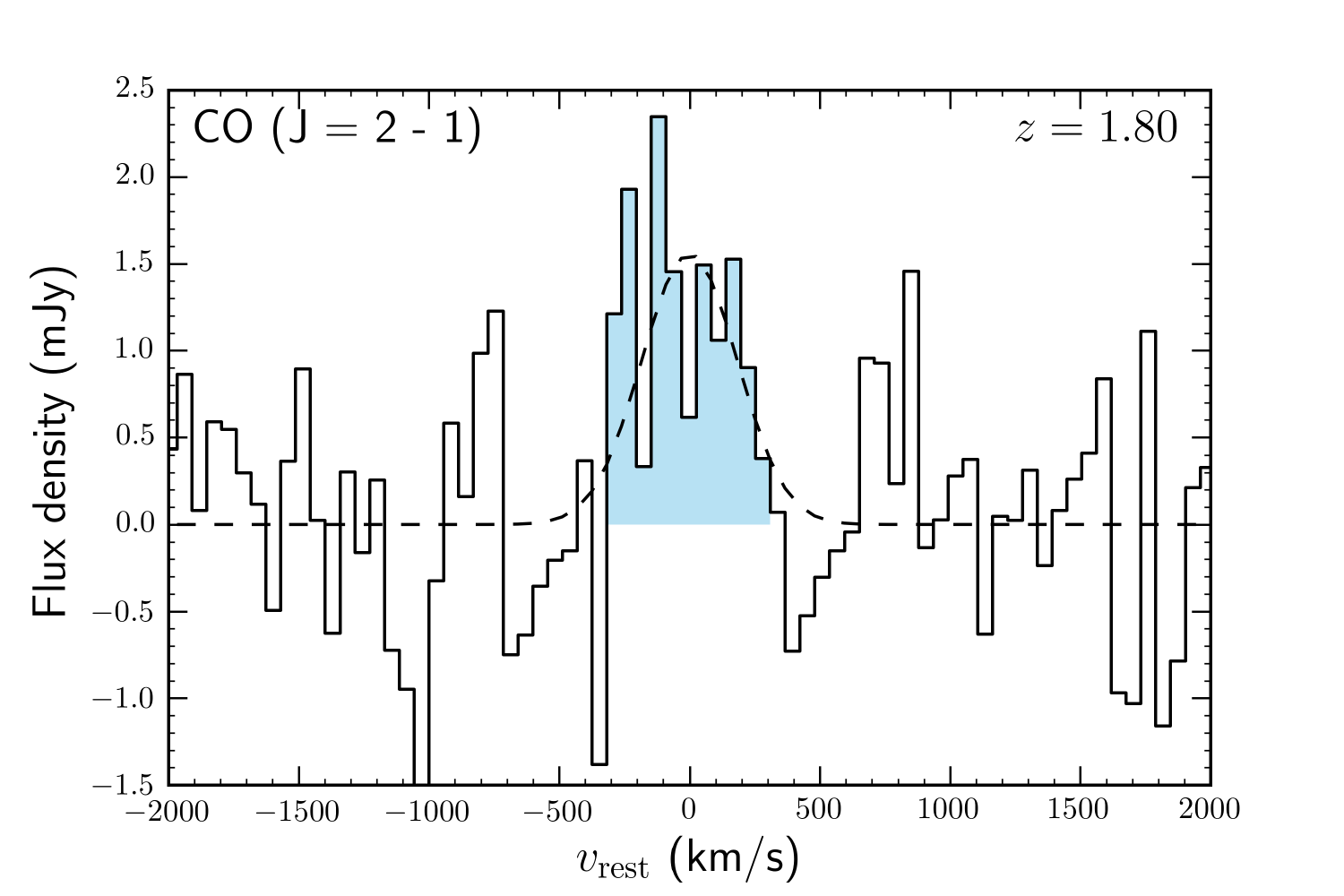}
	\caption{\emph{Left:} Moment-0 map for CO($J = 2 \to 1$) emission from J141917+524922, a serendipitously detected ULIRG at $z = 1.8$; the moment-0 image was made by integrating over the velocity range $-295$ km s$^{-1}$ to $273$ km s$^{-1}$. The source is detected at $\sim 7\sigma$ significance, where $1\sigma \sim 0.07$ Jy km s$^{-1}$. The contours at $\pm 2,3,4,5$ and $6\sigma$ significance have been overlaid on the 3D-HST $H$ band image. The synthesized beam has a size of $\sim 6.0'' \times 5.0''$ and is indicated in the bottom left corner. The image is shown without primary beam correction. \emph{Right:} Spectral line profile extracted for J141917+524922 from the primary-beam corrected spectral cube, at the peak pixel in the moment-0 map, with a velocity resolution of $\sim 56$ km s$^{-1}$. The dashed line shows a single-component Gaussian fit to the spectral line.}
	\label{fig:egs141917plots}
\end{center}			
\end{figure*}

We detect a spectral feature at nearly $\sim 7\sigma$ significance at $\nu_{\rm obs} =  82.289$ GHz. We fit the spectrum with a 1-D Gaussian  (\autoref{fig:egs141917plots}). We find a FWHM velocity width of $\Delta v_{\rm FWHM}  = 481 \pm 112 $ km s$^{-1}$ (\autoref{tab:table2}). We fit a 2-D Gaussian to the moment-0 map for our detected spectral feature to obtain the peak emission position (J2000 RA: 14h19m17.406s, Dec: 52d49m21.155s), which corresponds to a distance of $23.9''$ from the phase center of our observations. Based on the AEGIS source catalog and \emph{Spitzer} IRAC and optical counterparts in the HST-CANDELS field, we identify EGS J141917.4+524922 as the nearest counterpart (hereafter J141917+524922). This source is classified as a starburst-dominated ULIRG in the literature, with a known redshift of $z = 1.80\pm 0.02$ based on previously obtained IRS spectroscopy \citep{huang2009}. The detected emission thus is consistent with CO($ J = 2 \to 1$) emission at $z = 1.8016 \pm 0.0002$. Comparing our emission peak position to the optical $H$-band position for J141917+524922, we find a spatial offset of $\sim 0.46'' \pm 0.41''$, which is much smaller than our beam size ($6.0'' \times 5.0''$). We obtain a CO velocity-integrated line flux of $I_{\rm CO} = 0.75 \pm 0.11$ Jy km s$^{-1}$, after correcting for the primary beam response (\autoref{tab:table2}). 

\subsubsection{EGSIRAC J141912.03+524924.0}

\begin{figure*}
	\begin{center}
	\includegraphics[width=0.40\textwidth]{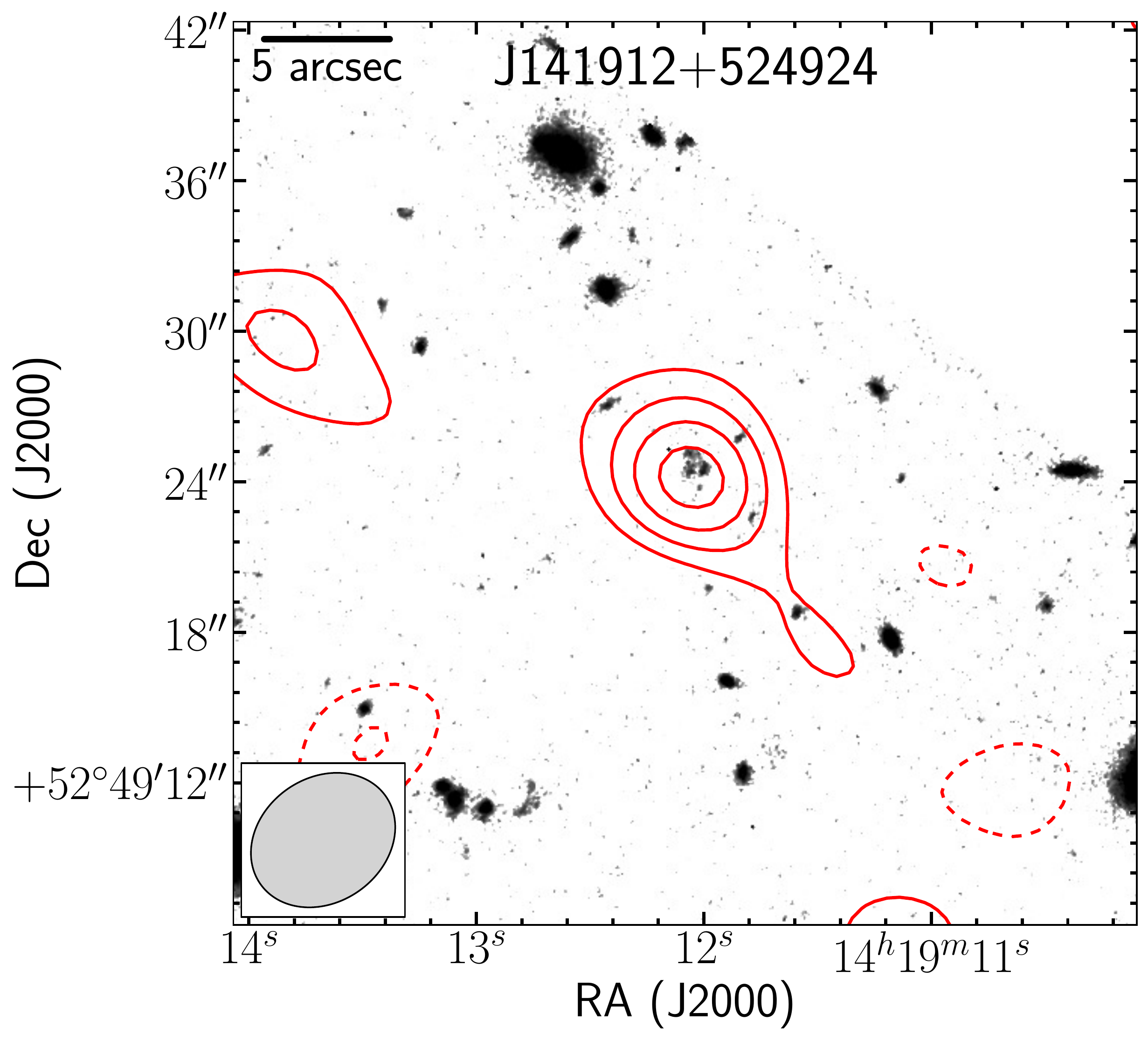}
	\includegraphics[width=0.59\textwidth]{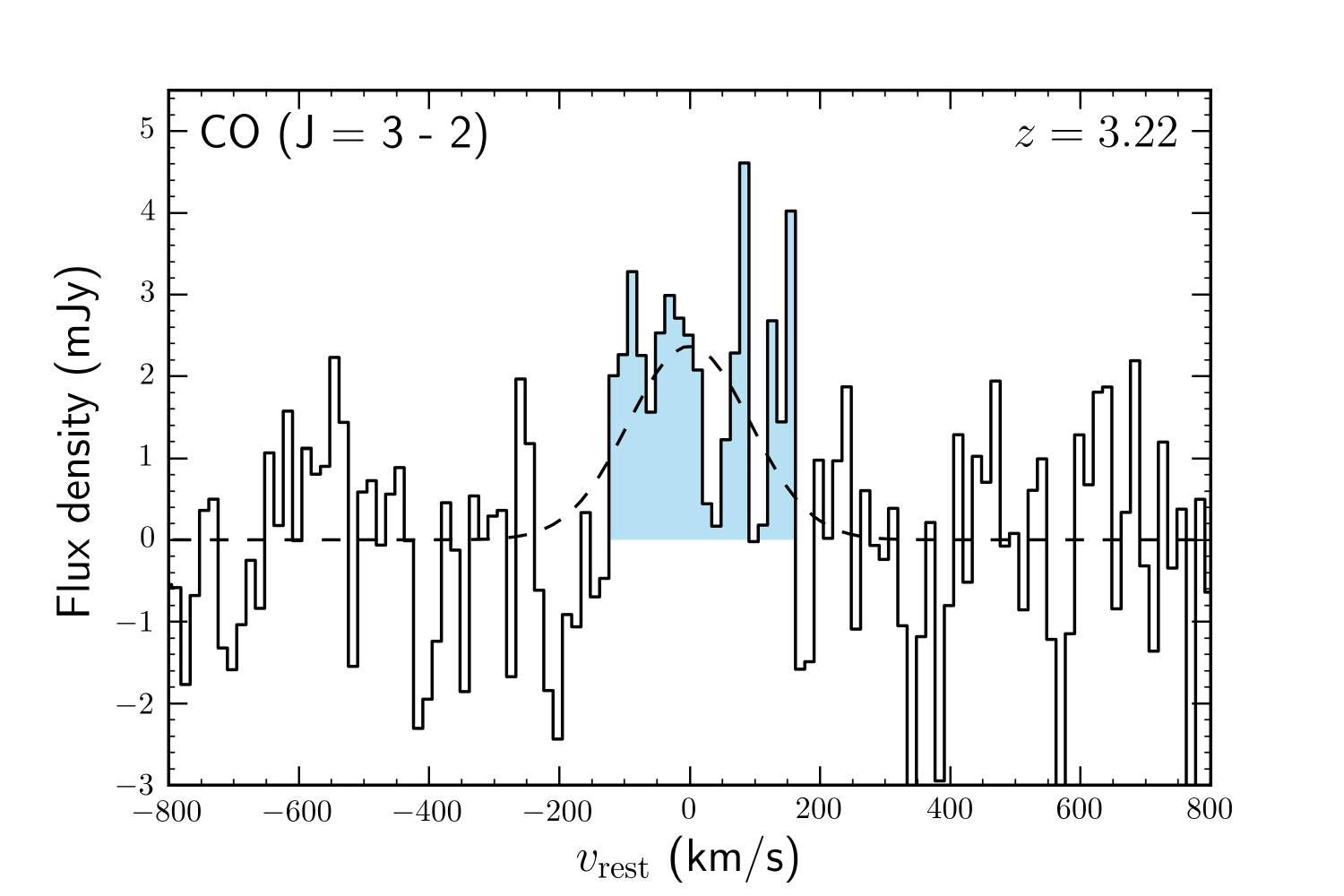}
	\caption{\emph{Left:} Moment-0 map for line emission from J141912+524924, a MS galaxy at $z = 3.22$. The line has been tentatively identified as CO($J = 3 \to 2$). The moment-0 image was made by integrating over the velocity range $-138$ km s$^{-1}$ to $134$ km s$^{-1}$. The source is detected at $\sim 6\sigma$ significance, where $1\sigma \sim 0.05$ Jy km s$^{-1}$. The contours at $\pm 2,3,4,5$ and $6\sigma$ significance have been overlaid on the 3D-HST $H$ band image. Both the $z_{\rm phot}$ and our detected spectral feature are consistent with CO($J = 3 \to 2$) line emission at $z = 3.22$.  The synthesized beam has a size of $\sim 6.0'' \times 5.0''$ and is indicated in the bottom left corner. The image is shown without primary beam correction. \emph{Right:} Spectral line profile extracted for J141912+524924 from the primary-beam corrected spectral cube, at the peak pixel in the moment-0 map, with a velocity resolution of $\sim 14$ km s$^{-1}$. The dashed line shows a single-component Gaussian fit to the spectral line.}
	\label{fig:egs141912plots}
	\end{center}	
\end{figure*}

We detect a spectral feature at $\sim 6.5\sigma$ significance at $\nu_{\rm obs}  = 81.930$ GHz. We fit the spectrum with a 1-D Gaussian (\autoref{fig:egs141912plots}). We find a FWHM velocity width of $\Delta v_{\rm FWHM} \sim 233 \pm 54$ km s$^{-1}$ (\autoref{tab:table2}). We fit a 2-D Gaussian to the moment-0 map for our detected spectral feature to obtain the peak emission position (J2000 RA: 14h19m12.088s, Dec: 52d49m24.235s), which corresponds to a distance of $27.0''$ from the phase center of our observations. Based on the AEGIS and the 3D-HST/CANDELS catalogs for the AEGIS field, we identify EGSIRAC J141912.03+524924.0 as the nearest counterpart (hereafter J141912+524924). Comparing the emission peak position to the optical $H$-band position for J141912+524924, we find an offset of $\sim 0.57'' \pm 0.55''$, which is much smaller than our beam size ($6.0'' \times 5.0''$). 

Unfortunately, there is no known redshift for this source to allow for immediate spectroscopic confirmation, although it has photometric coverage as part of the AEGIS catalog \citep{whitaker2011}. We therefore use the photometric redshift code EAZY \citep{brammer2008} to estimate a photometric redshift of $z_{\rm phot} \sim 3.23$. We test the robustness of this estimate using a variety of different population synthesis models, and find $2.6 < z_{\rm phot} < 4.2$. The emission feature thus is most consistent with the CO($J = 3 \to 2$) line at $z = 3.2206 \pm 0.0002$.  We obtain a velocity-integrated line flux of $I_{\rm CO} = 0.59 \pm 0.09$ Jy km s$^{-1}$, after correcting for the primary beam response (\autoref{tab:table2}). 

\subsection{SED fitting}\label{sec:sedfitting}

We perform SED fitting for all our detected sources using both CIGALE (Code Investigating GALaxy Emission; \citealt{noll2009, serra2011}) and the high-$z$ extension of MAGPHYS (Multi-wavelength Analysis of Galaxy Physical Properties; \citealt{dacunha2008,dacunha2015}). We use all available photometric data points for each of the sources, as described in \autoref{sec:arch}. \emph{Herschel}/PACS and \emph{Herschel}/SPIRE fluxes from the NEWFIRM Medium-band Survey (NMBS) are available only for EGS 13004291 and J141917+524921, as J141912+524924 is not detected. We used the de-blended \emph{Herschel} fluxes from the NMBS survey \citep{whitaker2011}. Since the two serendipitously detected sources are at distances of $23''$ and $27''$ from EGS 13004291 respectively, and can potentially contaminate its SPIRE fluxes (SPIRE beam size at 350 \mum$ $ $ \sim 24''$), the Herschel Interactive Processing Environment (HIPE) software package was used to examine the photometric images. The HIPE tasks SourceExtractorTimeline and source extractors were used to ensure that the de-blended FIR fluxes were not contaminated. 

CIGALE builds galaxy SEDs from UV to radio wavelengths assuming a combination of modules. These allow us to model the star formation history (SFH), the stellar emission using population synthesis models \citep{bruzual2003, maraston2005}, nebular lines, dust attenuation \citep[e.g.,][]{calzetti2000}, dust emission \citep[e.g.,][]{draine2007, casey2012}, contribution from an Active Galactic Nucleus (AGN; \citealt{dale2014,fritz2006}), and radio emission. The SEDs are built while maintaining consistency between UV dust attenuation, and FIR emission from the dust. We use simple analytical functions to model the star formation histories - a double exponentially decreasing SFH, and a delayed SFH. We use the dust attenuation from \cite{calzetti2000}, and the dust emission models from \cite{dale2014}. Finally, CIGALE performs a probability distribution function analysis for our specified model parameters, and obtains the likelihood-weighted mean value for each.  

MAGPHYS similarly uses a Bayesian approach to constrain galaxy-wide physical properties, including the star formation rate, stellar and dust mass, and dust temperature. It builds a large library of reference spectra with different star formation histories (using stellar population synthesis models from \citealt{bruzual2003}) and dust attenuation properties (using models from \citealt{charlot2000}). It also ensures energy balance between the optical and UV extinction and the FIR emission due to dust. 

In order to account for possible AGN contamination in our FIR-luminous sources, we carry out SED fitting with both CIGALE and MAGPHYS for EGS 13004291 and J141917+524922. EGS 13004291 has been detected in the $0.5 - 2$ keV soft X-ray band in deep \emph{Chandra} observations of the AEGIS field \citep{laird2009,nandra2015}, raising the possibility that an AGN may be present, and that it may contribute to its FIR luminosity. However, we can rule out the presence of an AGN for two reasons. First, the galaxy is consistent with the FIR-X-ray correlation \citep{syme2011}. Second, the hardness ratio of the X-ray emission indicates a starburst origin of the X-ray emission rather than an AGN \citep{syme2014}. We further test for the presence of an AGN in both our sources by including the fractional contribution of an AGN to the IR luminosity as a free parameter in SED fitting with CIGALE; no good fit is found with a non-zero AGN contribution. 

\begin{figure*}
	\includegraphics[width=\textwidth]{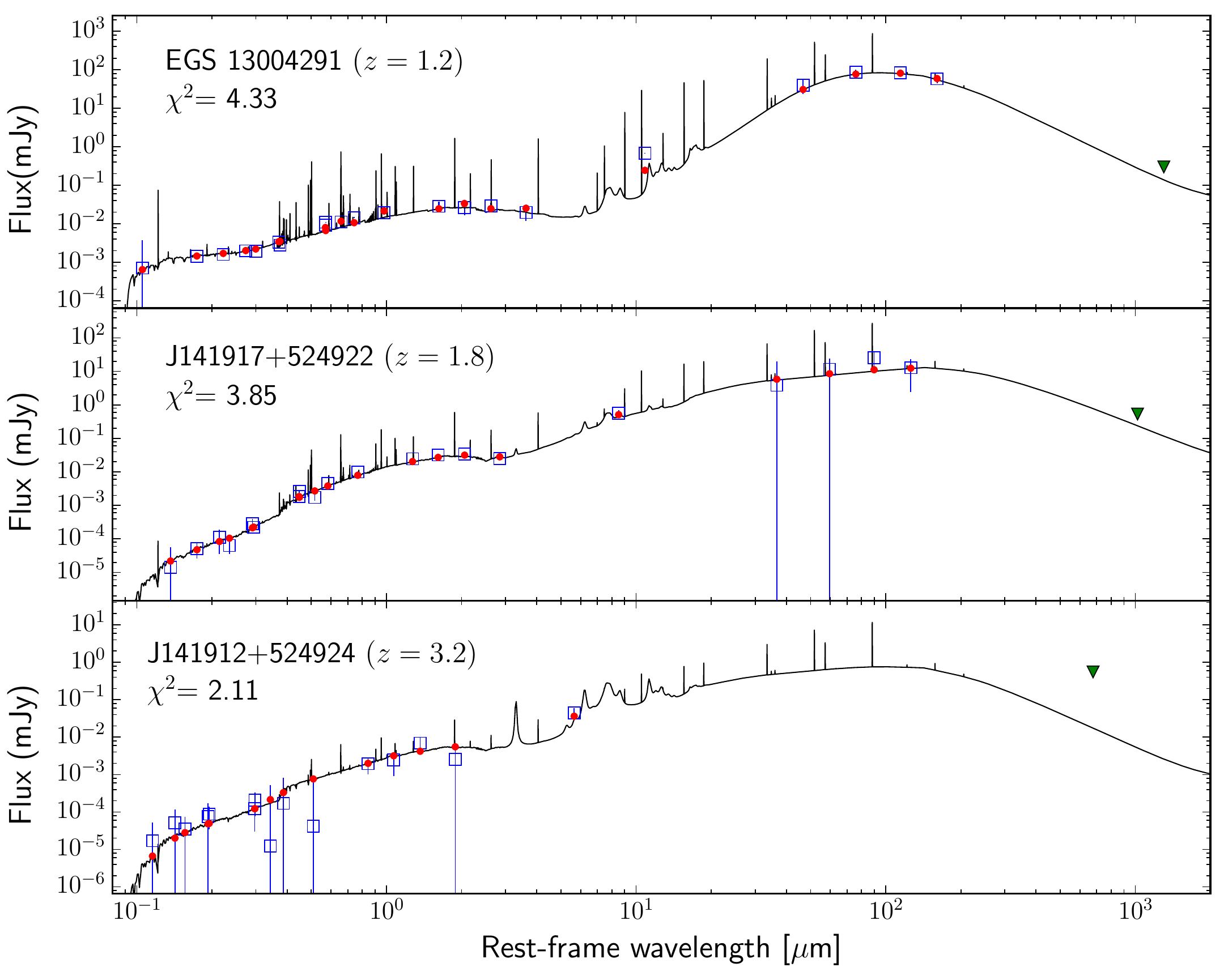}
	\caption{Best-fit SEDs (black lines) for EGS 13004291 (\emph{top}), J141917+524922 (\emph{middle}) and J141912+524924 (\emph{bottom}) derived using CIGALE. The blue squares represent the archival photometry up to an observed frame wavelength of 500 \mum, the red points represent the predicted model luminosities at the same wavelengths. The green triangles represent our new $3\sigma$ upper limits derived from the 3 mm continuum observations.}
	\label{fig:egs1412915_17_seds}
\end{figure*}

SED fits found using MAGPHYS and CIGALE were consistent within the errors for EGS 13004291 and J141912+524924. We did not find a good fit for the dust peak for J141917+524922 using MAGPHYS. The best-fit SEDs from CIGALE are shown in \autoref{fig:egs1412915_17_seds}, and the corresponding fit parameters are listed in \autoref{tab:table3}. 
\lfir$ $ for all sources was obtained by integrating the area under the best-fit SED over $\lambda_{\rm rest} = 42.5 - 122.5$ \mum, and $L_{\rm IR} $ by integrating over $\lambda_{\rm rest} = 8 - 1000$ \mum$ $ \citep{casey2014}. 

\begin{table*}[]
	\centering 
	\caption{\textsc{SED-fitting results and other physical parameters}}
	\begin{tabular}{l|c|c|c|c|c|c|c|c|c}
		\hline
		\hline
		Source				& $z$ 		& 	\lfir 		& 	$M_{\rm dust}$			& 	$M_{\rm gas}$ 			& $M_{*}$  					& SFR 						& sSFR				&  $\delta_{\rm GDR}$ 	& $f_{\rm gas}$\\
							& 			& ($10^{11} L_{\odot}$)	& (10$^{9}$ $M_{\odot}$) 	& (10$^{11}$ $M_{\odot}$)	& (10$^{11}$ $M_{\odot}$)	&( $M_{\odot}$ yr$^{-1}$)	& (Gyr$^{-1}$)		& 						& 				\\
		\hline
		EGS 13004291 	&$1.197$ 	&	$12.57 \pm 4.2 $ 			& 	$2.4 \pm 0.2$			& 	$2.8 \pm 0.3$  			&  $1.0 \pm 0.1$ 			& $ 714 \pm 42$  			& $6.8 \pm 0.7$ 	& $117\pm 21 $ 		& $0.74$ 		 \\

		J141917+524922	&$1.802$ 	&  	$3.15 \pm 1.1 $ 			& 	$3.7 \pm 3.6$ 			& 	$1.5 \pm 0.2$  			&  $2.5 \pm 0.2$ 			& $ 384 \pm 40$  			& $1.6 \pm 0.5$ 	& $40 \pm 44$  		& $0.37$  		\\
		J141912+524924  	&$3.221$ 	&	$1.69 \pm 0.5 $	    			& 	$0.7 \pm 0.5$ 			& 	$2.6 \pm 0.4$  			&  $0.5 \pm 0.1$ 			& $ 110 \pm 7$   			& $2.2 \pm 0.3$ 	& $380 \pm  342$ 		& $0.83$		 \\ 
		\hline \noalign {\smallskip}
	\end{tabular}
	
	{\textbf{Notes:} \lfir$ $ is calculated by integrating the best-fit SED model over $\lambda_{\rm rest} = 42.5 - 122.5$\mum. The gas masses were calculated using $\alpha_{\rm CO} =  3.6 M_{\odot} $(K km s$^{-1}$  pc$^{2})^{-1}$ (see \autoref{sec:consistencyMS}) and the luminosities listed in \autoref{tab:table1}, assuming brightness temperature ratios of $r_{21} = 0.76$ and $r_{32} = 0.42$.} 
	
	\label{tab:table3} 
\end{table*}

\subsection{Comparison with literature values}\label{sec:comparison}

We compare our SED-fitting derived physical parameters (\autoref{tab:table3}) against those available in the literature. For EGS 13004291, \cite{tacconi2013} have derived a star formation rate of SFR $\sim (630 \pm 220)$ $M_{\odot} $yr$^{-1}$ and a stellar mass of $M_{*} \sim (9.3 \pm 2.8) \times 10^{10} M_{\odot}$. These values were obtained using an extinction-corrected 24\mum$ $ derived IR luminosity + UV flux. However, \cite{freundlich2013} derive a SFR $\sim 182$ $M_{\odot}$ yr$^{-1}$, using the optical [O {\sc ii}] line luminosity, and a stellar mass of $M_{*} \sim 5.0 \times 10^{11} M_{\odot}$. They add the caveat that this method is likely to underestimate the SFR, since it does not account for dust-embedded star-forming regions. 

Our derived properties from CIGALE depend significantly on the assumed star formation history for EGS 13004291. Assuming a stellar history including a young stellar population from recent starburst activity, we find a SFR $\sim 714 \pm 42 $ $M_{\odot} $yr$^{-1}$, averaged over the last 10 Myr, and a stellar mass of  $M_{*} \sim 1.0 \times 10^{11} M_{\odot}$. Using only the \lfir$ $ as a star formation indicator \citep{kennicutt1998}, we obtain a SFR $\sim 220$ $M_{\odot}$ yr$^{-1}$, which is comparable to the SFR obtained using UV emission; the sum of the SFRs obtained from UV and FIR emission is then roughly consistent with the SFR from SED fitting.

For EGS 141917+524922, previous SED fitting has been performed to the IR and radio photometry \citep{huang2009}, on which basis it was classified as a starburst dominated ULIRG. Our stellar mass of $M_{*} \sim (2.5 \pm 0.2)\times 10^{11}$ is consistent with the literature value of $M_{*} \sim 3 \times 10^{11}$. However, we find a $L_{\rm IR}$ $\sim (9 \pm 3) \times 10^{11} L_{\odot}$, which is $5 \times $ lower than the literature value. We expect the uncertainties  on our estimate to be smaller due to better constraints on the IR luminosity provided by the \emph{Herschel}/SPIRE photometry, which samples the peak of the FIR emission. Using only the \lfir, we obtain an SFR $\sim 96$ $M_{\odot}$ yr$^{-1}$, while we obtain a SFR $\sim (384 \pm 40) M_{\odot}$ yr$^{-1}$ from SED fitting;  as the SED fitting incorporates both the UV and FIR photometry, this implies that the dust-obscured star formation plays a relatively minor role.  

Similarly for EGS 141912+524924, we find a SFR $ \sim 30$ $M_{\odot}$ yr$^{-1}$ using only the \lfir; this is significantly smaller than the SFR estimated using complete optical photometric data SFR $(\sim 110 \pm 7)$ $M_{\odot}$ yr$^{-1}$, indicating that the most of the star formation is not dust-obscured. \newline 

\section{Discussion}\label{sec:discussion}

\subsection{CO excitation in EGS 13004291}\label{sec:coex}

The quiescent mode of gas consumption in SFGs at high redshifts can result in markedly different molecular gas excitation than in the ULIRG/SMG/QSO population at comparable redshifts. While some SMGs show a low excitation molecular gas component \citep[e.g.,][]{riechers2011b,ivison2011}, and although there is significant scatter in their excitation properties \citep[e.g.,][]{sharon2016}, the CO excitation of the brightest ULIRGs and SMGs can be nearly thermalized up to $J = 3$, and mid-$J$ CO lines can be used to trace the bulk of the molecular gas, which is in a warm dense medium in many of these galaxies \citep[e.g.,][]{riechers2006b, riechers2009a, riechers2013}. However, the bulk of the molecular gas in MS galaxies lies in extended, cold, diffuse gas reservoirs, and can be most reliably traced using low-excitation molecular gas lines, as shown by the subthermal CO excitation prevalent in BzK galaxies (massive, optically selected SFGs at high-$z$ ; see \citealt{dannerbauer2009, daddi2010a, daddi2015}). For a limited sample of BzK galaxies, there is also evidence for the presence of an additional warm and dense molecular gas component, resulting in significant CO($J = 5 \to 4$) emission, which has been suggested to result from giant dense starbursting clumps \citep{freundlich2013, bournaud2015, daddi2015}. However, the observed warm, dense molecular gas component traced by high-$J$ CO lines does not encompass the cold, dense molecular gas traced by low-$J$ HCN emission, which is critical for star formation. 

We measure the velocity integrated line intensity for CO($J = 2 \rightarrow 1 $) to be $I_{\mathrm{CO}(J = 2 \to 1)} =  3.09 \pm 0.27$ Jy km s$^{-1}$, resulting in a line luminosity of $L'_{\mathrm{CO}(J = 2 \to 1)} = (6.0 \pm 0.5) \times 10^{10}$ K km s$^{-1}$ pc$^{2}$ (\autoref{fig:co21}). Based on the $L'_{\mathrm{CO}(J = 3 \to 2)}$ value in the literature \citep{tacconi2013}, we obtain a brightness temperature ratio of $r_{32}$ $\sim 0.65 \pm 0.15$. This is consistent with the average $r_{32} = 0.58 \pm 0.16$ for \hiz BzK galaxies \citep{aravena2014,daddi2015}, all of which show significantly subthermal molecular gas excitation at $J \geq 3$. This is however in sharp contrast to \hiz ULIRGs such as the Cloverleaf quasar and F10214+4724, which display a molecular gas excitation consistent with a single high temperature medium i.e. the CO rotational lines are thermalized to high-$J$ values ($r_{32} \sim 1$;  \citealt{riechers2011}). 

The conversion factor $\alpha_{\rm CO}$ used to derive the molecular gas mass from the CO line luminosity generally depends on the metallicity, the average gas density, the relative fractions of warm and cold dense gas, and the gas excitation in each source \citep[see][for reviews]{solomon2005, carilli2013, bolatto2013}, and different values are appropriate for ULIRGs as opposed to normal SFGs. Here, we attempt to determine the most appropriate $\alpha_{\rm CO}$ for EGS 13004291, based on its dynamical mass.  We obtain an estimate for the dynamical mass based on our observed FWHM velocity width  $\Delta v_{\rm FWHM}  = 340 \pm 40$ km s$^{-1}$ and the previously observed CO($ J = 3\to2$) half-light radius $r_{1/2} = 3.9 \pm 1.0$ kpc, as determined by \cite{tacconi2013}.  In addition, we consider the source morphology and orientation. The position-velocity (PV) diagrams for EGS 13004291, using both optical [O {\sc ii}] and CO spectral line observations, appear to show a disk-like velocity profile  \citep{freundlich2013}. Assuming a disk-like structure for EGS 13004291, we use a 2-D Gaussian fitting to the optical \emph{HST} $H$-band image to find an axial ratio of $b/a \sim 0.7$ i.e. sin$^{2}(i) \sim 0.5$.  We then assume a dynamical mass model for a rotating disk, which can be described by $M_{\rm dyn}$ sin$^{2}(i) = 233.5$ $r_{1/2} (\Delta v_{\rm FWHM})^{2}$, where $\Delta v_{\rm FWHM}$ is in km s$^{-1}$  and $r_{1/2}$ is the half-light radius of the molecular disk in pc \citep{solomon2005}. We thus obtain a dynamical mass of $M_{\rm dyn} = (21 \pm 15) \times 10^{10} M_{\odot}$. 

Based on the relation between the dynamical mass within the half-light radius, gas mass, and the assumed dark matter (DM) mass,  $M_{\rm dyn} = 0.5( M_{*} + M_{\rm gas}) + M_{\rm DM}$ \citep{daddi2010b}, we then calculate the gas mass to be $M_{\rm gas} = (22 \pm 15) \times 10^{10} M_{\odot}$  (assuming that DM contributes 25\% to the dynamical mass). This implies a CO luminosity to H$_{2}$ gas mass conversion factor of $\alpha_{\rm CO} \sim (2.8 \pm 2.0)$ $M_{\odot} $(K km s$^{-1}$ pc$^{2})^{-1}$,  consistent with the $\alpha_{\rm CO} = 3.6$ $M_{\odot} $(K km s$^{-1}$ pc$^{2})^{-1}$ typically used for \hiz MS galaxies \citep[e.g.][]{daddi2015}. 

However, we caution that the use of the dynamical mass to estimate $\alpha_{\rm CO}$ suffers from systematic uncertainties resulting from the dynamical estimator used. For example, if we assume an isotropic virial estimator \citep[e.g.][]{pettini2001, binney2008, engel2010}

\begin{equation} 
M_{\rm dyn} (r < r_{1/2}) = 190  r_{1/2} (\Delta v_{\rm FWHM})^{2}
\end{equation}

using a normalization for a rotating disk at an average inclination such that sin$^{2}( i$)$ = 0.25$ \citep[see][]{engel2010,bothwell2010}, we obtain a dynamical mass of $M_{\rm dyn} = (8.5 \pm 6.1) \times 10^{10} M_{\odot}$ within $r_{1/2} \sim 3.9$ kpc, which implies a  gas mass of  $M_{\rm gas} = (2.8 \pm 2.0) \times 10^{10} M_{\odot}$ and a conversion factor $\alpha_{\rm CO} = (0.4 \pm 0.3) $ $M_{\odot} $(K km s$^{-1}$ pc$^{2})^{-1}$. This is more consistent with the values typically found in SMGs and nearby ULIRGs, $\alpha_{\rm CO} = 0.8 M_{\odot} $(K km s$^{-1}$ pc$^{2})^{-1}$  \citep[e.g.][]{downes1998,hodge2012,riechers2014}. Adjusting the above calculation to our inclination such that sin$^{2}( i$)$ = 0.5$ would further decrease the $M_{\rm dyn}$ by a factor of $\times 2$, which would be inconsistent with our stellar mass  $M_{*} \sim 1.0 \times 10^{11} M_{\odot}$.  This discrepancy in $M_{\rm dyn}$ estimates highlights the systematic uncertainties inherent to dynamical mass computations.  Motivated by the constraints on the dust to gas mass ratio as discussed in Section \autoref{sec:consistencyMS}, we adopt $\alpha_{\rm CO} = 3.6$ $M_{\odot} $(K km s$^{-1}$ pc$^{2})^{-1}$. This is consistent with the $\alpha_{\rm CO}$ derived for PHIBSS galaxies (including EGS 13004291) by \citet{carleton2016} under the assumption of a constant gas depletion timescale. 

\subsection{Derived properties from SED fitting}\label{sec:consistencyMS}

We here use the SFR, $M_{*}$, $M_{\rm dust}$ from the SED fitting 	(see \autoref{tab:table3}) , and the $M_{\rm gas}$ determined from our CO observations, to compare our sources to the galaxy MS at their respective redshifts. We use the sSFR/sSFR$_{\rm MS}$, the gas-to-dust mass ratio $\delta_{\rm GDR}$, and the gas fraction $f_{\rm gas}$ as diagnostics, where sSFR$_{\rm MS}$ is the predicted sSFR for the galaxy MS at the source redshift \citep{whitaker2012}, $ \delta_{\rm GDR} = M_{\rm gas}/ M_{\rm dust}$, and $f_{\rm gas}$ is defined as 

\begin{equation}
f_{\rm gas} = \frac{M_{\rm gas}}{M_{\rm gas} + M_{*}}.
\end{equation}

For EGS 13004291, we find a sSFR/sSFR$_{\rm MS}$ $\sim 13 \pm 3 $, consistent with its classification as a starburst galaxy \citep{tacconi2013,genzel2015}. To estimate the dust temperature,  we approximate the \cite{draine2007} dust model by a blackbody multiplied by a power-law opacity and find a dust temperature of $T_{\rm dust} \approx 20 U_{\rm min}^{0.15} \sim 30$ K, where $U_{\rm min}$ is the best-fit intensity of the radiation field from SED fitting \citep{aniano2012}. Assuming an $\alpha_{\rm CO} =  3.6 M_{\odot} $(K km s$^{-1}$ pc$^{2})^{-1}$, we find a gas mass of $M_{\rm gas} = (2.8 \pm 0.3 ) \times 10^{11} $ M$_{\odot}$ and a gas fraction of $f_{\rm gas} \sim 0.74$, consistent with the literature value ($f_{\rm gas}  = 0.79; $ \citealt{tacconi2013}). We also find a gas-to-dust mass ratio of $\delta_{\rm GDR} \sim 120$.  $\delta_{\rm GDR}$ strongly depends on the metallicity of the system \citep[e.g.,][]{leroy2011, sandstrom2013, remy2014, groves2015}, and can vary significantly depending on the assumed dust properties, although typical values for \hiz galaxies are close to $\delta_{\rm GDR} \sim 100$ \citep{casey2014}. For the $z \sim1.5$ MS galaxy BzK21000, a $\delta_{\rm GDR} \sim 104$ was found \citep{magdis2011}. In low and high metallicity MS galaxies at $z \sim 1.4$,  \cite{seko2016} find $\delta_{\rm GDR} \sim 570$ and $\delta_{\rm GDR} \sim 400$, respectively. However, their derived $\delta_{\rm GDR}$ may be lower by a factor of $2-3\times$ depending on the dust model assumed. Thus, the $\delta_{\rm GDR}$ in EGS 13004291 is consistent with that seen in $z \sim 1-2$ gas-rich galaxies.  Alternatively, if we were to assume an $\alpha_{\rm CO} =  0.8 M_{\odot} $(K km s$^{-1}$ pc$^{2})^{-1}$, we would obtain a gas mass of $M_{\rm gas} = (6 \pm 1 ) \times 10^{10} $ M$_{\odot}$, a gas fraction of $f_{\rm gas} \sim 0.38$, and a gas-to-dust mass ratio of $\delta_{\rm GDR} = 26 \pm 21$, which is more consistent with the observed gas to dust mass ratio $\delta_{\rm GDR}\sim 40$ for dusty SMGs at high-$z$ \citep[e.g.,][]{santini2010, bothwell2013, swinbank2014,zavala2015}. $\alpha_{\rm CO}$ varies inversely with the CO excitation temperature $T_{\rm ex}$ \citep{bolatto2013} i.e. lower $\alpha_{\rm CO}$ values are more typically seen in sources such as SMGs and quasars, which also show higher CO excitation \citep{ivison2011,carilli2013}, while EGS 13004291 displays subthermal CO excitation consistent with other MS galaxies at $z \sim 1.5$ \citep{daddi2015}. Therefore, we finally adopt an $\alpha_{\rm CO} =  3.6 M_{\odot} $(K km s$^{-1}$ pc$^{2})^{-1}$ for our $M_{\rm gas}$ and $f_{\rm gas}$ calculations (see \autoref{tab:table3}). 

For J141917+524922, we find a sSFR/sSFR$_{\rm MS}$ $\sim 2.5$, which is consistent with the galaxy MS at $z = 1.8$ within the scatter \citep{genzel2015}. This motivates our choice of $\alpha_{\rm CO} =  3.6 M_{\odot} $(K km s$^{-1}$ pc$^{2})^{-1}$, instead of the value typically used for ULIRGs, i.e., $\alpha_{\rm CO} =  0.8 M_{\odot} $(K km s$^{-1}$ pc$^{2})^{-1}$. Similarly, using a brightness temperature ratio $r_{21} = 0.76$ \citep{daddi2015}, we obtain a gas mass of $M_{\rm gas} = (1.5 \pm 0.2) \times 10^{11} M_{\odot}$ and a gas fraction of $f_{\rm gas} = 0.37 $, which is again consistent with the galaxy MS at $z =1.8$. We obtain a gas-to-dust mass ratio of $ \delta_{\rm GDR} = 40 \pm 44$, which is consistent with the average value found for SMGs. The large uncertainties on the FIR fluxes of J141917+524922 result in a poorly constrained dust mass and therefore gas-to-dust ratio. 

Applying similar diagnostics to J141912+524924, we find a sSFR/sSFR$_{\rm MS}$ $\sim 0.8$, making this source consistent with the galaxy MS at $z = 3.22$. J141912+524924 is then the highest redshift unlensed MS galaxy detected to date. We find a gas-to-dust ratio $ \delta_{\rm GDR}  = 380 \pm 342$, and a gas fraction of $f_{\rm gas} \sim 0.83$, implying that it is a gas-rich galaxy with possibly low dust content. We emphasize that the lack of FIR photometry for this source leads to a poorly constrained dust mass, and hence gas-to-dust ratio.

\subsection{Comparison of dense gas properties}\label{sec:compdense}

We use the following set of diagnostics to characterize star formation in EGS 13004291: the fraction of dense, actively star-forming gas  $ f_{\rm dense} = M_{\rm dense}/M_{\rm gas} \propto$ \lhcn/\lco, the global star formation efficiency of the molecular gas SFE$_{\rm mol}  =  {\rm SFR}/M_{\rm gas}  \propto $ \lfir/\lco$ $, and the star formation efficiency of the dense gas SFE$_{\rm dense} =   {\rm SFR}/M_{\rm dense}  \propto$ \lfir/\lhcn$ $. Unless explicitly stated, we calculate these ratios using line luminosities for the $J = 1 \to 0$ transition, calculated in \autoref{sec:gobs},\autoref{sec:dgobs}. We calculate \lfir/\lhcn, \lhcn/\lco, and \lfir/\lco$ $ values for the sample of archival sources described in \autoref{sec:arch}, and compare our obtained values for EGS 13004291 against those for different galaxy populations using these diagnostics. Our sample of archival sources includes the Cloverleaf quasar and IRAS F10214+4724, which are among the best studied and brightest HCN and CO sources at high-$z$. This allows us to place the properties of its star-forming environment in context. We do not compare our source to APM 08279+5255 at $z = 3.9$, the only other solid HCN detection at high-$z$; we have excluded this source despite detailed observations because of its unusual gas excitation properties \citep[e.g.,][]{weiss2007,riechers2010b}. 

\begin{figure}
\centering
	\includegraphics[width=0.44\textwidth]{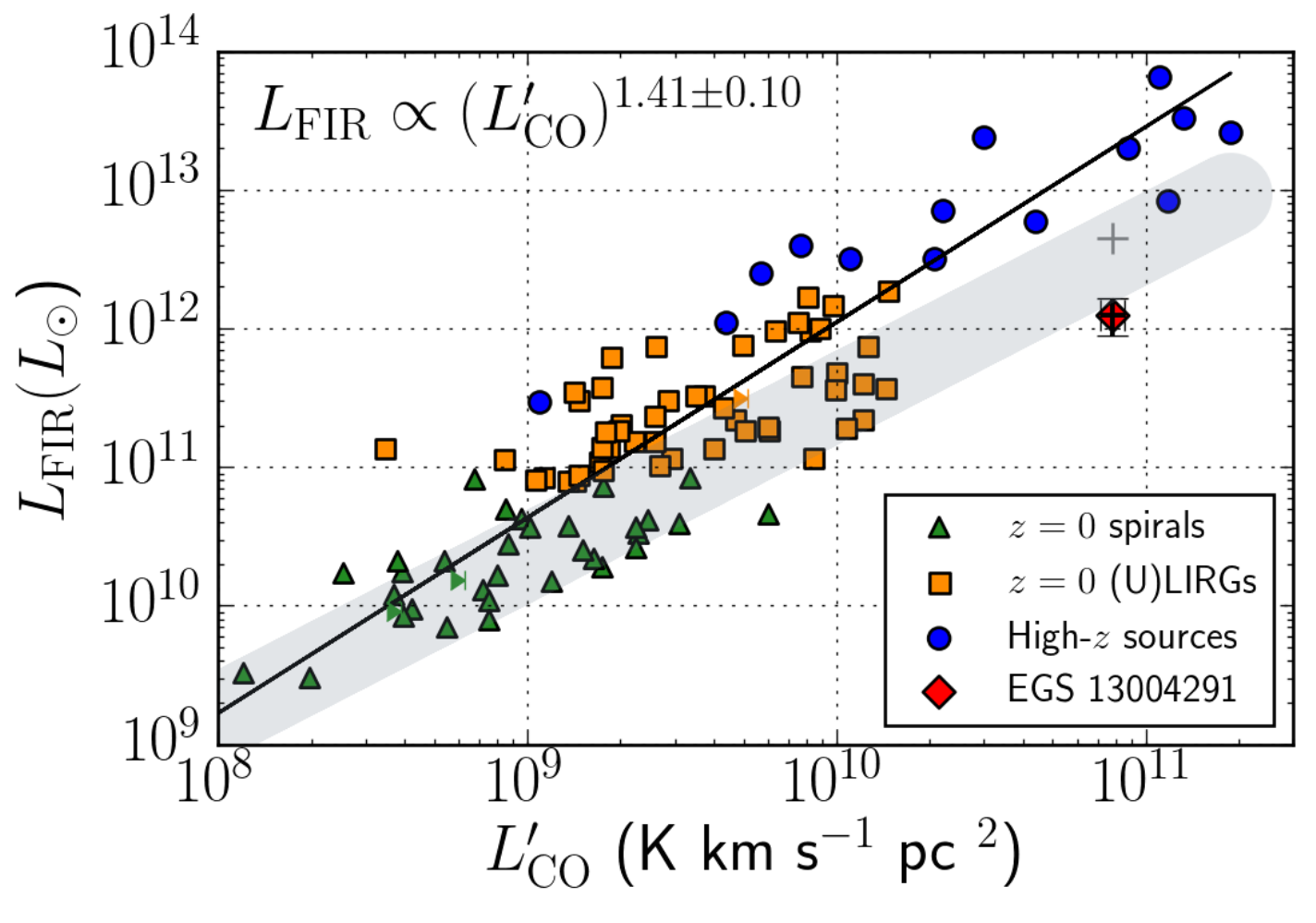}
	\includegraphics[width=0.44\textwidth]{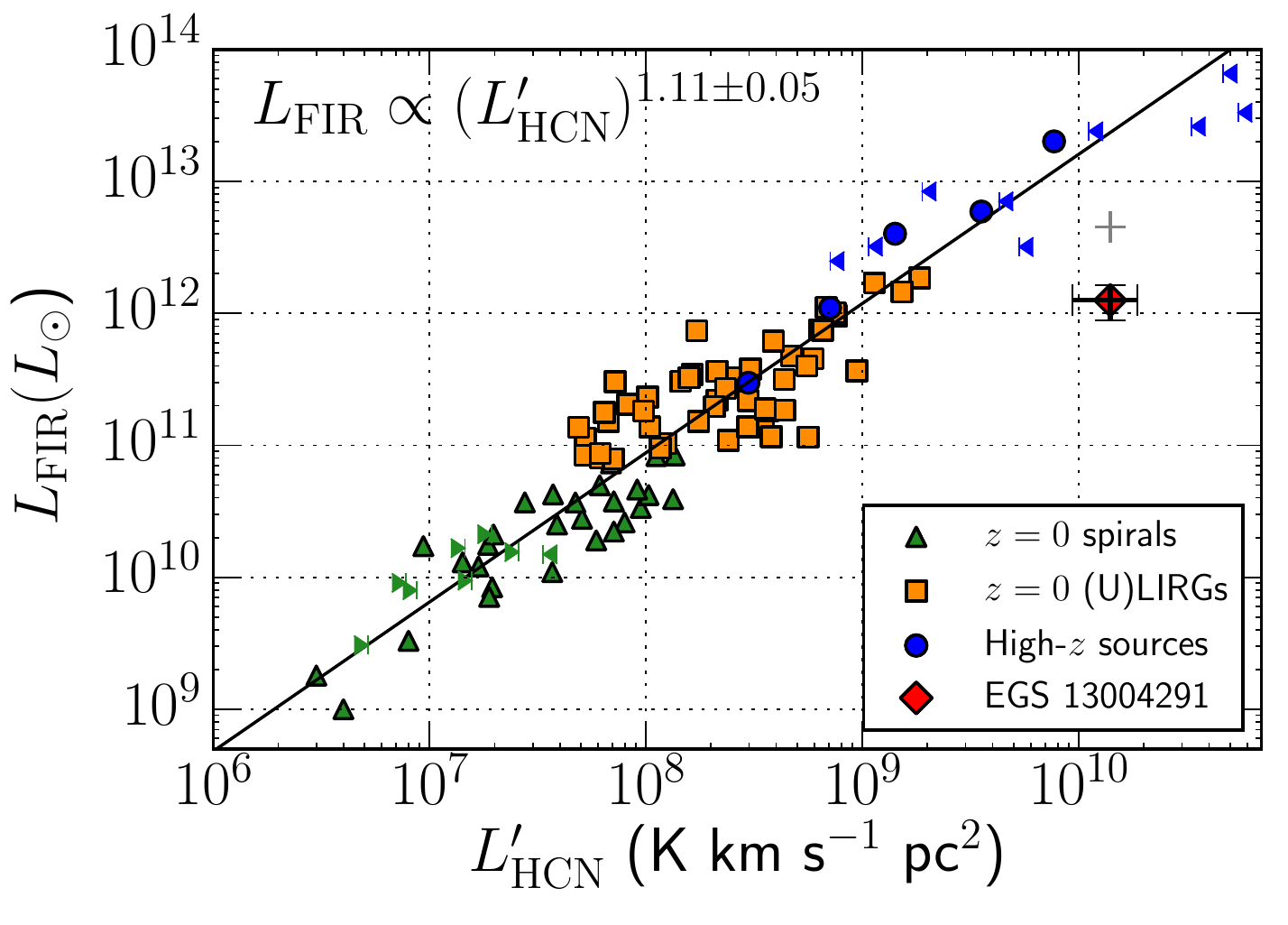}
	\includegraphics[width=0.46\textwidth]{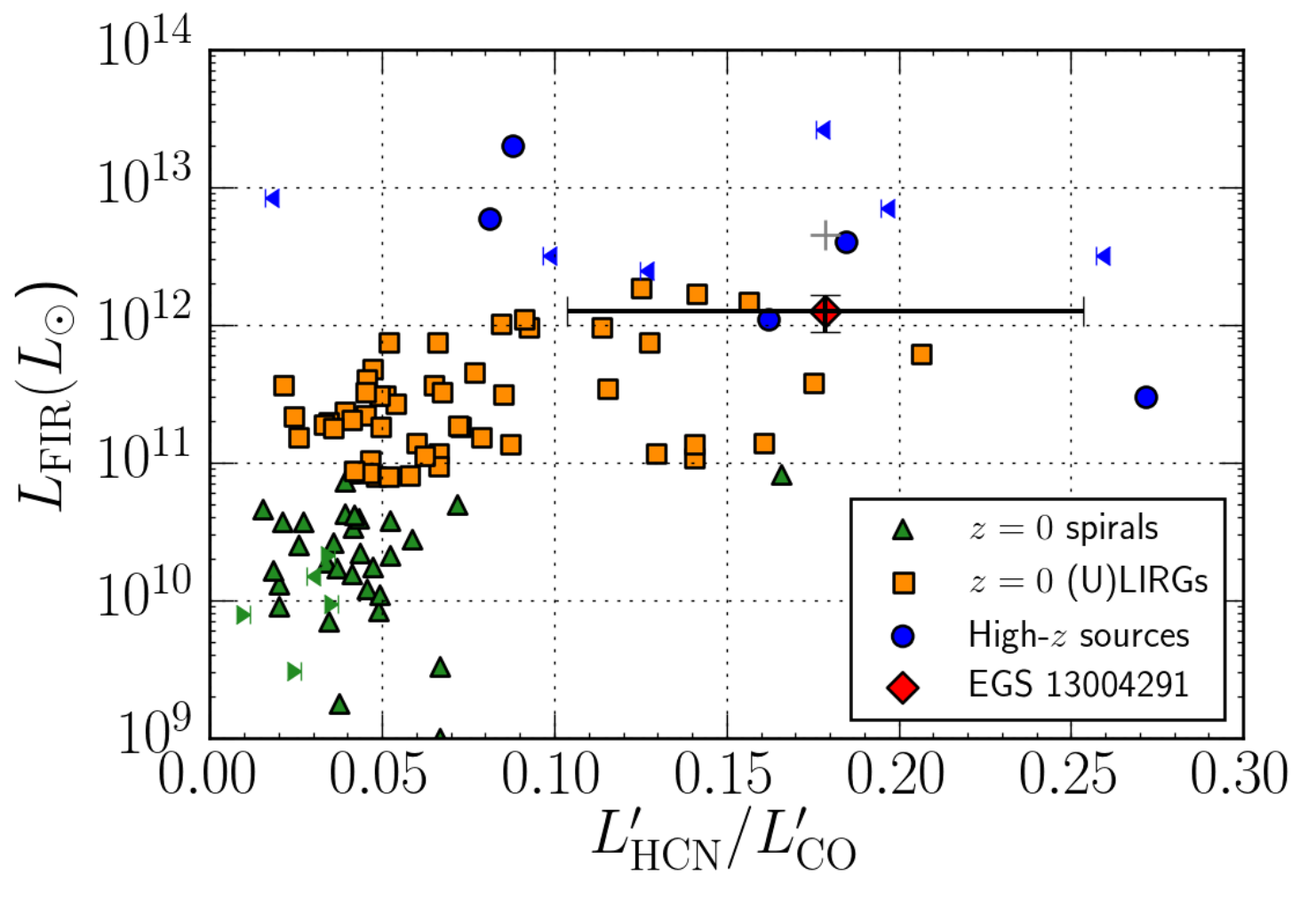}

	\caption{\lfir-\lco$ $ (\emph{top}), \lfir-\lhcn$ $ (\emph{middle})  and \lfir -$f_{\rm dense}$ (\emph{bottom}) relations for the compiled archival data on a large sample of sources for which CO and HCN rotational lines have been observed, and where IR photometric coverage exists, as described in \autoref{sec:arch}. The values have been updated to the assumed cosmology; the green upper triangles represent the local spiral galaxy population \citep{gao2004b}, the orange squares represent the updated values for (U)LIRGs \citep{gracia2008, garcia2012}, and the blue circles represent all $z > 1$ sources \citep{gao2007, riechers2007}; the left or right pointed triangles represent upper or lower limits on detections; the shaded region in the top plot shows the $\pm 1\sigma$ region around the best-fit to the galaxy main-sequence \citep{genzel2010}. The best fit lines were derived using only the detections. The red diamond represents EGS 13004291, which lies below the best fit line in \lfir-\lhcn, implying that EGS 13004291 has a \emph{lower} SFE$_{\rm dense}$ when compared to this galaxy sample. The grey cross represents the source EGS 13004291, with an \emph{equivalent} \lfir$ $ which represents the total star formation, instead of only the dust-obscured star formation, enabling better comparison against the \hiz sources in our sample whose star formation is predominantly dust-obscured (see \autoref{sec:eqfir}).} 
	\label{fig:lfir-lhcn-lco}
\end{figure}

\subsubsection{Star formation models}\label{sec:sfm}

Two main classes of dense gas star formation models have been proposed to explain the observed \lfir-\lhcn$ $ relation (see \autoref{fig:lfir-lhcn-lco}, middle): threshold density models, and turbulence-regulated star formation models. Threshold density models assume a fixed SFR per unit gas mass above a certain critical density, i.e. a constant SFE$_{\rm dense}$. Working under this assumption, the slope of the \lfir-\lhcn$ $ relation ($\equiv$ SFE$_{\rm dense}$) is expected to remain constant, regardless of the star-forming environment or the galaxy type \citep{gao2004b, wu2005, wu2010}. 

On the other hand, turbulence-regulated star formation models predict that the SFR depends on the properties of the molecular gas cloud as a whole - the gas surface density, pressure, and the turbulent velocity dispersion \citep[e.g.,][]{krumholz2005, krumholz2007b}. Based on these properties, they predict that different molecular gas tracers would produce different slopes in the \lfir-gas luminosity plot. In their parametrization, SFE$_{\rm mol}$ is a function of both the local free-fall timescale, $\tau_{\rm ff}$, and of the ISM turbulence represented by the Mach number $\mathcal{M}$. At a given $\mathcal{M}$, the SFE varies according to $1/\tau_{\rm ff}$, and the SFR varies according to $\bar{n}/\tau_{\rm ff}$, where  $\bar{n}$ is the average gas density. The observed SFE$_{\rm mol}$ then depends on how the critical density ($n_{\rm crit}$) of the tracer in question compares to $\bar{n}$, as this determines the local free-fall timescale, i.e. $\tau_{\rm ff}$ will be smaller and SFE$_{\rm mol}$ will be higher in regions of large density contrasts. For CO observations, which trace all molecular gas, the average gas density is comparable to the critical density and $\tau_{\rm ff} \propto \bar{n}^{-0.5}$, and the SFR $\propto \bar{n}^{1.5}$, which is close to the slope seen in the \lfir-\lco$ $ relation. If the average gas density is much smaller than the tracer's critical density, $\tau_{\rm ff}$ is determined by $n_{\rm crit}$, and the SFE will be nearly constant i.e the slope of the \lfir-gas luminosity relation is close to unity, thus explaining the slope of the \lfir-\lhcn$ $ relation. Additionally, the SFE varies inversely with the Mach number $\mathcal{M}$ i.e. increased turbulence in the ISM hinders efficient star formation. 

In this model, $f_{\rm dense}$ increases with both the average gas density $\bar{n}$ and ISM turbulence, as a higher Mach number $\mathcal{M}$ broadens the density probability distribution of $f_{\rm dense}$, and hence increases the mass of the dense gas above the critical density. Finally, SFE$_{\rm mol}$ $= f_{\rm dense} \times \rm {SFE_{\rm dense}}$; we can thus constrain SFE$_{\rm dense}$ using this model. 

Observationally, the turbulence-regulated star formation scenario is favored. Merging ULIRGs at both \hiz and in the local Universe display a higher \lfir/\lhcn, which is consistent with the prediction that SFE$_{\rm dense}$ increases with $\bar{n}$ \citep[e.g.,][]{riechers2007, gracia2008,bussmann2008,juneau2009, garcia2012}. Spatially resolved studies of the star-forming regions in local galaxies show that \lfir/\lhcn$ $ is strongly dependent on environment, and \emph{anti-correlated} with \lhcn/\lco \citep[e.g.,][]{ usero2015,bigiel2016}. \lfir/\lhcn$ $ is lower in galaxy nuclear regions, which typically show the highest \lhcn/\lco, and increases in regions of low \lhcn/\lco; the overall \lco/\lfir$ $ remains nearly constant. This has been interpreted as the increase in SFE$_{\rm dense}$ in regions of large density contrasts, instead of regions with large dense gas fractions, which naturally leads to a lower SFE$_{\rm dense}$ in galaxy nuclei, despite a large $f_{\rm dense}$ \citep{bigiel2016}. All these observational results are in conflict with the idea that star formation proceeds with a universal constant efficiency above a certain threshold density. 

An additional source of uncertainty is the HCN to dense gas mass conversion factor $\alpha_{\rm HCN}$, such that $M_{\rm dense} =  \alpha_{\rm HCN} L'_{\mathrm{HCN}(J = 1 \to 0)}$. Studies in the literature suggest that the same $\alpha_{\rm HCN}$ is not applicable to mergers or extreme galaxies \citep[e.g.,][]{garcia2012}, and that the $\alpha_{\rm HCN}$ in ULIRGs needs to be lower by a factor similar to $\alpha_{\rm CO}$, relative to the MW canonical values. This has been taken into consideration while comparing the threshold density and turbulence-regulated star formation models \citep{usero2015}. \cite{garcia2012} also find that threshold density models place much more stringent constraints on the allowed $\alpha_{\rm HCN}$ and $\alpha_{\rm CO}$ values, as opposed to turbulence-regulated models.

\subsubsection{Dense molecular gas properties of EGS 13004291}\label{sec:ahcn}

Using our observed HCN emission, and the \lfir$ $ obtained from SED fitting, we calculate \lfir/\lhcn$ $ for EGS 13004291. Comparing our obtained value against the \lfir/\lhcn$ $ values for our compiled sample of galaxies (as described in \autoref{sec:arch}), we find that it lies below the best-fit \lfir-\lhcn$ $ relation (\autoref{fig:lfir-lhcn-lco}, middle). Thus, its SFE$_{\rm dense}$ (traced by \lfir/\lhcn) is lower than that seen in other galaxy populations. This is inconsistent with the concept of dense gas clumps as fixed units of star formation having a constant SFE$_{\rm dense}$. 

EGS 13004291 has a line luminosity ratio of \lhcn/\lco $= 0.17 \pm 0.07$, significantly higher than that seen in local spiral galaxies (\lhcn/\lco $= 0.04 \pm 0.02$), but consistent with that seen in local ULIRGs (\lhcn/\lco $= 0.13 \pm 0.03$) and \hiz (U)LIRGs (\lhcn/\lco $= 0.16 \pm 0.07$). Given that this line luminosity ratio traces the dense gas fraction $f_{\rm dense}$ (modulo $\alpha_{\rm HCN}$/$\alpha_{\rm CO})$, this implies that the dense gas fraction in EGS 13004291 is significantly higher than observed in normal local galaxies, and is instead consistent with that displayed by some of the most extreme \hiz sources, including the Cloverleaf quasar and F10214+4724 \citep{solomon2003,vanden2004,solomon2005}.

We can rule out the possibility that the enhanced dense gas fraction we find is caused by $\alpha_{\rm HCN}$ or by our assumed $r_{21}$ to our results as follows. We have assumed $r_{21} = 0.65$ for the HCN excitation, which is the mean value as observed in local starburst/AGN dominated galaxies \citep{krips2008, geach2012}. While no similar studies have been performed for normal local spiral galaxies, they are likely to have lower $r_{21}$ values. The maximum values for pure starburst galaxies are $r_{21} \lesssim 0.7$ \citep{krips2008}, while the most extreme values have been seen upon zooming into the nuclei of galaxies with nuclear starbursts  (e.g. $r_{21} \sim 1$ in the nuclear region of M82, \citealt{krips2008}). Such extreme conditions ($T_{k} \sim 40$ K, with the CO being partially optically thin, see \citealt{turner1990}) are unlikely to prevail over large spatial scales in EGS 13004291, given the subthermal excitation between the CO($J  = 2 \to 1$) and CO($J  = 3 \to 2$) lines \citep{papa2012}. We therefore expect that the $r_{21}$ for EGS 13004291 is comparable to or less than $r_{21} \sim 0.65$. We therefore derive conservative estimates for the dense gas fraction, as lower values for $r_{21}$ will result in a higher $L'_{\mathrm{HCN}(J = 1 \to 0)}$ line luminosity and a larger dense gas fraction.

Assuming a galactic $\alpha^{\rm MW}_{\rm HCN} = 10 M_{\odot}$ (K km s$^{-1}$ pc$^{2})^{-1}$ \citep{gao2004b}  we obtain a dense gas mass of $M_{\rm dense} = (1.4 \pm 0.5) \times 10^{11} M_{\odot}$, which would imply that about half the gas mass is in the form of dense gas i.e, $f_{\rm dense} \sim 0.48$, whereas typical values for the MW and ULIRGs are typically $\sim 3\%$ \citep{lada2012} and $\sim 10\%$  \citep{gao2001}. This suggests that $\alpha_{\rm HCN}$ is likely lower in EGS 13004291 than assumed here. Mechanically driven turbulence or shocks, caused by infall or outflow of large amounts of molecular gas can also result in enhancement of HCN emission  and drive down the value of $\alpha_{\rm HCN}$ \citep{imanishi2014,martin2015}.

However, if the HCN luminosity is not enhanced relative to the other dense gas tracers observed in this system, we can extend the turbulence-regulated star formation model to explain the observed properties for EGS 13004291. It is possible that the observed SFE$_{\rm dense}$ is low because of a turbulent and compressive i.e. high-pressure ISM, undergoing constant replenishment by infalling gas from the cosmic web and being rapidly consumed by star formation. An increase in $\mathcal{M}$ results in a higher $f_{\rm dense}$ and a lower SFE$_{\rm dense}$, which would be consistent with our observations. This fits in with the picture of a near-constant SFE$_{\rm mol}$ in MS galaxies and starburst galaxies lying above the MS. \cite{silverman2015} find evidence that starburst galaxies lying above the galaxy MS show slightly higher SFE$_{\rm CO}$, but by a smaller factor than in local ULIRGs and starburst galaxies. In the framework described above, the SFE$_{\rm mol}$ depends on SFE$_{\rm dense}$ and $f_{\rm dense}$ which are more fundamental quantities; assuming that EGS 13004291 is not an exceptional source, this suggests that there can be naturally occurring variations in the SFE of starburst galaxies. In that case, the SFE$_{\rm mol}$ is not an adequate representative of star-formation properties in these galaxies, and dense gas observations will be even more critical for such sources.    

\subsubsection{Other dense gas tracers}

We obtain a brightness temperature ratio of \lhcn/\lhnc$ $ $\sim1.8 \pm 1.3$, and we obtain a \emph{lower} limit of \lhcn/\lhcop $\gtrsim 1.2$ for the $ J = 2 \to 1$ transitions. These ratios depend on multiple factors, including the relative abundance of HCN, HCO$^{+}$ and HNC and their different excitation states, given the gas density and temperature, and the gas ionization state  \citep{aalto1994,huett1995,lepp1996,meijerink2007}. Higher temperatures favor an increased HCO$^{+}$ abundance; conversely, the HNC abundance is favored over that of HCN at lower temperatures \citep{schilke1992, graninger2014}, leading to the observed inverse correlation between \lhcn/\lhcop$ $ and \lhcn/\lhnc$ $ \citep[e.g.][]{loenen2007}.

Galaxies hosting AGN typically show a higher \lhcn/\lhcop$ $ brightness temperature ratio than starburst or composite galaxies \citep[e.g.][]{kohno2001,imanishi2007,krips2008, izumi2015}, due to a relatively enhanced HCN abundance. \citet{privon2015} find a mean line ratio of \lhcn/\lhcop $\sim1.84$ for AGN hosts, while \lhcn/\lhcop  $\sim 1.14$ is found in AGN-starburst composite galaxies, and \lhcn/\lhcop $\sim 0.88$ is found in purely starburst galaxies. However, a high \lhcn/\lhcop$ $ ratio is not uniquely associated with the presence of an AGN, as this ratio is not solely driven by relative HCN and HCO$^{+}$ abundances. High \lhcn/\lhcop$ $ ratios can also be observed in UV-dominated Photon Dominated Regions (PDRs) with high densities, as well as in low-density X-ray Dominated Regions (XDRs; \citealt{meijerink2007,privon2015}). We assume that the \lhcn/\lhcop$ $ ratio in the $J = 2 \to 1$ lines is representative of that in the $J = 1 \to 0$ lines. This is a reasonable assumption for galaxies which are not AGN-dominated \citep[see][]{krips2008}. 

Our lower limit on \lhcn/\lhcop$\gtrsim 1.2$ is consistent with the line ratios observed in starburst galaxies  \citep{privon2015}, though higher than the mean line ratio (\lhcn/\lhcop $\sim 0.88$). Our limit is also consistent with the observed line ratio in the Cloverleaf quasar, HCN/HCO$^{+}$ $\sim 1.3$ \citep{riechers2006a}, where it is proposed to arise from optically thick emission from HCN and HCO$^{+}$ occupying regions of similar density and volume.

\lhcn/\lhnc$ $ is a sensitive probe of the gas temperature \citep{graninger2014},  and \lhcn/\lhnc$ $ has been observed to vary between \lhcn/\lhnc$ $ $\sim0.16$  to \lhcn/\lhnc$ $ $\sim 2$ within a similar class of galaxies (i.e. with similar dense gas fractions traced by \lhcn/\lco; \citealt{aalto2002}). We find \lhcn/\lhnc$ $ $\sim 1.8 \pm 1.3$, which is consistent but slightly higher than the canonical value for \lhcn/\lhnc$ $ $\sim 1.1$ \citep{goldsmith1981} in the absence of a highly-ionized medium.This is consistent with the expected increase in \lhcn/\lhnc$ $ in star forming regions \citep{meier2005}. The presence of shocks or mechanically driven turbulence would also contribute to a high \lhcn/\lhnc, as HNC is preferentially destroyed in shocks \citep{schilke1992,lindberg2016}. 

\subsubsection{Sources of uncertainty}\label{sec:eqfir}

We note our results are subject to the uncertainties inherent to the use of \lfir$ $ as a tracer for the star formation.  While a significant fraction of the star formation in nearby spirals and \hiz disk galaxies is not dust-obscured (see \autoref{sec:consistencyMS})  sources such as nearby ULIRGs, \hiz SMGs and FIR-luminous quasars are typically dominated by dust-obscured star formation. Thus, the \lfir$ $ is a reasonable estimator for the SFR in our archival sample of \hiz sources, all of which belong to the latter category. However, only 30\% of the star formation activity in EGS 13004291 is dust-obscured. We investigate the impact of this finding on our study by defining an ``equivalent" FIR luminosity $L_{\rm FIR}^{\rm eq}$, such that the total SFR obtained from our SED fitting (including both UV and FIR emission) corresponds to SFR$_{\rm UV + FIR} = 1.07 \times 10^{-10} L_{\rm FIR}^{\rm eq}$. The alternative constraints obtained using this value are shown as grey crosses in \autoref{fig:lfir-lhcn-lco} for comparison. We find that adopting $L_{\rm FIR}^{\rm eq}$ would not significantly affect our overall results.

\begin{table}[]
	\centering
	\caption{\textsc{Comparison of Source Properties}}
	\begin{tabular}{lccc}
		\hline
		\hline
		\vspace{1mm}
		Source  									& \lfir/\lhcn  					&  \lfir/\lco 					& \lhcn/\lco  			\\
		\vspace{1mm}
													&($L_{\odot}/L_{l})^{a}$		&	($L_{\odot}/L_{l})^{a}$	& 						\\
		\hline	
		\vspace{1mm}
		EGS 13004291  							& $90 \pm 30 $   		& $ 16 \pm 5 $   			& $0.17 \pm 0.07$         \\
		Other high-$z$ galaxies: 																									\\	
		\hspace{1mm} F10214+4724     			& $2600 \pm 550$   		& $ 523 \pm 130  $  		& $0.18 \pm 0.04$       	\\
		\hspace{1mm} Cloverleaf 					& $1300 \pm 300$ 			& $ 135 \pm 33 $ 			& $0.10 \pm 0.02$         \\
		Averages :																												 	\\
		\hspace{1mm} \hiz (U)LIRGs (5) 			& $1667 \pm 688$			& $251 \pm 129$ 			& $0.16 \pm 0.07$ 		\\

		\hspace{1mm} $z\approx0$ ULIRGs (5) 	& $1347 \pm 264 $			& $148 \pm 33$ 			& $0.13 \pm 0.03$ 		\\
		\hspace{1mm} $z\approx0$ LIRGs (85) 	& $829 \pm 791 $			& $40 \pm 70$ 				& $0.05 \pm 0.04$ 		\\
		\hspace{1mm} $z\approx0$ spirals (45)		& $618 \pm 395$			& $23 \pm 23$ 				& $0.04 \pm 0.02$		\\ 	
		\hline   \noalign {\smallskip}
	\end{tabular}
	\textbf{Notes:} $^{a} L_{l} = $ K km s$^{-1}$ pc$^{2}$; average values calculated from the archival data as described in \autoref{sec:arch}. 
	\label{tab:table4}
\end{table}

\subsection{Comparison with other MS galaxies}\label{sec:compMS}

A number of studies have focused on the bimodality of star formation modes in \hiz galaxies, with the bulk proceeding quiescently and a small population of starbursts with higher star formation efficiencies, and correspondingly shorter gas depletion timescales \citep[e.g.,][]{daddi2010b, rodighiero2011}. However, recent work on large samples of MS galaxies, and on outliers above the MS may suggest that all SFGs have the same gas depletion timescale $\tau_{\rm depl}$ and therefore the same star formation efficiency \citep[e.g.,][]{scoville2015}. These authors argue that this short, constant depletion time for both MS and above-MS galaxies is due to a different mode of SF prevailing at high redshifts, either driven by compressive and rapid gas motions and/or galaxy-galaxy mergers, with the dispersive gas motion arising from the rapid accretion needed to maintain the high SFR. This would be a more efficient mode for star formation from existing gas supplies at high-$z$, applying to both MS and above MS galaxies. 

EGS 13004291 lies above the MS, with a sSFR/sSFR$_{\rm MS}$ $\sim 13$. Despite being a starburst, we find that its SFE is not enhanced relative to other MS galaxies i.e. is consistent within the scatter \citep[see Fig. 6a, by][]{dz2015}. This fits in with the model described above, where both MS and above MS galaxies have the same gas depletion timescale. The lack of enhancement in the SFE$_{\rm mol}$ is also consistent with turbulence-regulated star formation models, since the relative effects of an enhanced $f_{\rm dense}$ and a lower SFE$_{\rm dense}$ cancel each other out. Given that our SFE$_{\rm mol}$ is consistent with that found in MS galaxies at the same epoch, our higher dense gas fraction also fits in with a star formation mode where the ISM is compressive and more turbulent; star formation would no longer proceed in isolated dense gas clumps as in the Milky Way, but in a more widespread manner throughout the disk, so the idea of dense gas clumps as discrete units of star formation would no longer be applicable. This is similar to the prevalent picture in ULIRGs \citep{solomon2005}, although in a milder form because it is driven by less extreme processes than merging galaxies. 

\section{Conclusions}\label{sec:conclusions}

We have tentatively detected dense molecular gas, using HCN and HNC as tracers, and constrained the CO excitation in EGS 13004291, one of the most gas-rich star-forming galaxies detected at high-$z$. We find the following: 

\begin{itemize}

	\item The CO excitation in EGS 13004291 is subthermal and consistent with that seen in BzK galaxies, and significantly different from IR-luminous sources such as the Cloverleaf quasar and F10214+4724, which show nearly thermalized CO emission up to $J = 4$.
	\item The SFE in EGS 13004291 is consistent with literature values for the galaxy MS at $z \sim 1$, unlike starbursts in the local Universe which show higher SFEs as compared to normal galaxies.
	\item The dense gas fraction $f_{\rm dense}$ for EGS 13004291, as traced by \lhcn/\lco$ = 0.17 \pm 0.07$, is significantly enhanced over that in normal spiral galaxies at $z \sim 0$, and but is instead consistent with the values typically displayed by ULIRGs and those found for SMGs at high-$z$.
	\item Additionally, we report the  serendipitous detection of CO rotational transitions in two \hiz sources in our HCN observations, one of which, if confirmed, would be the highest redshift MS galaxy detected in CO to date. 

\end{itemize} 

Overall, we find that the SFE for the starburst EGS 13004291 is consistent with that for normal MS galaxies at high-$z$, which is in sharp contrast to the difference found between normal and starburst galaxy populations in the local Universe. Together with the enhanced dense gas fraction observed in EGS 13004291, this indicates that the rapid star formation in \hiz starburst galaxies does not proceed analogously to that in local starburst galaxies. More critically, this difference cannot be quantified by using only \lfir/\lco$ $ as a proxy for the SFE.  Our observations support the model of a constant depletion timescale in both MS and starburst galaxies, resulting from a turbulent, compressive ISM; the logical next step is to perform such studies for a larger sample of MS and starburst galaxies at high-$z$. Observations of the dense gas component will be vital, as studies of the \lfir/\lco$ $ alone will miss variations in the SFE$_{\rm dense}$ and $f_{\rm dense}$. Such observations are now feasible with both ALMA and NOEMA, and will demonstrate whether enhanced dense gas fractions are universally found in MS and starburst galaxies, or whether EGS 13004291 is an outlier, with an enhanced dense gas fraction resulting from an infall or outflow of molecular gas. Concurrent observations of other dense gas tracers are also important, to confirm the utility of HCN as a dense gas tracer, and are well within the reach of ALMA and NOEMA's wide bandwidth receivers. Finally, spatially resolved studies of dense gas star formation in the local Universe, such as the ongoing EMPIRE survey \citep{bigiel2016}, will help to clarify the role of turbulence-regulated or density threshold models in star formation.  

\acknowledgments
We would like to thank the anonymous referee for highly detailed and constructive comments that improved the clarity of our results. We thank Denis Burgarella, for his invaluable assistance with the CIGALE SED-fitting code, and Gordon Stacey for enlightening comments on the CO-to-H$_{2}$ conversion factor. This research has made use of NASA's Astrophysics Data System.DR and RP acknowledge support from the National Science Foundation under grant number AST-1614213 to Cornell University. R.P. acknowledges support through award SOSPA3-008 from the NRAO. H.D. acknowledges financial support from the Spanish Ministry of Economy and Competitiveness (MINECO) under the 2014 Ram\'on y Cajal program MINECO RYC-2014-15686. This study makes use of data from AEGIS, a multiwavelength sky survey conducted with the Chandra, GALEX, Hubble, Keck, CFHT, MMT, Subaru, Palomar, \emph{Spitzer}, VLA, and other telescopes and supported in part by the NSF, NASA, and the STFC. This work is based on observations taken by the 3D-HST Treasury Program (HST-GO-12177 and HST-GO-12328) with the NASA/ESA Hubble Space Telescope, which is operated by the Association of Universities for Research in Astronomy, Inc., under NASA contract NAS5-26555. This study makes use of data from the NEWFIRM Medium-Band Survey, a multi-wavelength survey conducted with the NEWFIRM instrument at the KPNO, supported in part by the NSF and NASA.

\clearpage

\clearpage


\begin{thebibliography}{}
	\expandafter\ifx\csname natexlab\endcsname\relax\def\natexlab#1{#1}\fi
	
	\bibitem[{{Aalto} {et~al.}(1994){Aalto}, {Booth}, {Black}, {Koribalski}, \&
		{Wielebinski}}]{aalto1994}
	{Aalto}, S., {Booth}, R.~S., {Black}, J.~H., {Koribalski}, B., \&
	{Wielebinski}, R. 1994, \aap, 286, 365
	
	\bibitem[{{Aalto} {et~al.}(2002){Aalto}, {Polatidis}, {H{\"u}ttemeister}, \&
		{Curran}}]{aalto2002}
	{Aalto}, S., {Polatidis}, A.~G., {H{\"u}ttemeister}, S., \& {Curran}, S.~J.
	2002, \aap, 381, 783
	
	\bibitem[{{Aniano} {et~al.}(2012){Aniano}, {Draine}, {Calzetti}, {Dale},
		{Engelbracht}, {Gordon}, {Hunt}, {Kennicutt}, {Krause}, {Leroy}, {Rix},
		{Roussel}, {Sandstrom}, {Sauvage}, {Walter}, {Armus}, {Bolatto}, {Crocker},
		{Donovan Meyer}, {Galametz}, {Helou}, {Hinz}, {Johnson}, {Koda}, {Montiel},
		{Murphy}, {Skibba}, {Smith}, \& {Wolfire}}]{aniano2012}
	{Aniano}, G., {Draine}, B.~T., {Calzetti}, D., {et~al.} 2012, \apj, 756, 138
	
	\bibitem[{{Aravena} {et~al.}(2014){Aravena}, {Hodge}, {Wagg}, {Carilli},
		{Daddi}, {Dannerbauer}, {Lentati}, {Riechers}, {Sargent}, \&
		{Walter}}]{aravena2014}
	{Aravena}, M., {Hodge}, J.~A., {Wagg}, J., {et~al.} 2014, \mnras, 442, 558
	
	\bibitem[{{Barmby} {et~al.}(2008){Barmby}, {Huang}, {Ashby}, {Eisenhardt},
		{Fazio}, {Willner}, \& {Wright}}]{barmby2008}
	{Barmby}, P., {Huang}, J.-S., {Ashby}, M.~L.~N., {et~al.} 2008, \apjs, 177, 431
	
	\bibitem[{{Bigiel} {et~al.}(2016){Bigiel}, {Leroy}, {Jim{\'e}nez-Donaire},
		{Pety}, {Usero}, {Cormier}, {Bolatto}, {Garcia-Burillo}, {Colombo},
		{Gonz{\'a}lez-Garc{\'{\i}}a}, {Hughes}, {Kepley}, {Kramer}, {Sandstrom},
		{Schinnerer}, {Schruba}, {Schuster}, {Tomicic}, \&
		{Zschaechner}}]{bigiel2016}
	{Bigiel}, F., {Leroy}, A.~K., {Jim{\'e}nez-Donaire}, M.~J., {et~al.} 2016,
	\apjl, 822, L26
	
	\bibitem[{{Binney} \& {Tremaine}(2008)}]{binney2008}
	{Binney}, J., \& {Tremaine}, S. 2008, {Galactic Dynamics: Second Edition}
	(Princeton University Press)
	
	\bibitem[{{Blain} {et~al.}(2002){Blain}, {Smail}, {Ivison}, {Kneib}, \&
		{Frayer}}]{blain2002}
	{Blain}, A.~W., {Smail}, I., {Ivison}, R.~J., {Kneib}, J.-P., \& {Frayer},
	D.~T. 2002, \physrep, 369, 111
	
	\bibitem[{{Bolatto} {et~al.}(2013){Bolatto}, {Wolfire}, \&
		{Leroy}}]{bolatto2013}
	{Bolatto}, A.~D., {Wolfire}, M., \& {Leroy}, A.~K. 2013, \araa, 51, 207
	
	\bibitem[{{Bothwell} {et~al.}(2010){Bothwell}, {Chapman}, {Tacconi}, {Smail},
		{Ivison}, {Casey}, {Bertoldi}, {Beswick}, {Biggs}, {Blain}, {Cox}, {Genzel},
		{Greve}, {Kennicutt}, {Muxlow}, {Neri}, \& {Omont}}]{bothwell2010}
	{Bothwell}, M.~S., {Chapman}, S.~C., {Tacconi}, L., {et~al.} 2010, \mnras, 405,
	219
	
	\bibitem[{{Bothwell} {et~al.}(2013){Bothwell}, {Smail}, {Chapman}, {Genzel},
		{Ivison}, {Tacconi}, {Alaghband-Zadeh}, {Bertoldi}, {Blain}, {Casey}, {Cox},
		{Greve}, {Lutz}, {Neri}, {Omont}, \& {Swinbank}}]{bothwell2013}
	{Bothwell}, M.~S., {Smail}, I., {Chapman}, S.~C., {et~al.} 2013, \mnras, 429,
	3047
	
	\bibitem[{{Bouch{\'e}} {et~al.}(2010){Bouch{\'e}}, {Dekel}, {Genzel}, {Genel},
		{Cresci}, {F{\"o}rster Schreiber}, {Shapiro}, {Davies}, \&
		{Tacconi}}]{bouche2010}
	{Bouch{\'e}}, N., {Dekel}, A., {Genzel}, R., {et~al.} 2010, \apj, 718, 1001
	
	\bibitem[{{Bournaud} {et~al.}(2015){Bournaud}, {Daddi}, {Wei{\ss}}, {Renaud},
		{Mastropietro}, \& {Teyssier}}]{bournaud2015}
	{Bournaud}, F., {Daddi}, E., {Wei{\ss}}, A., {et~al.} 2015, \aap, 575, A56
	
	\bibitem[{{Brammer} {et~al.}(2008){Brammer}, {van Dokkum}, \&
		{Coppi}}]{brammer2008}
	{Brammer}, G.~B., {van Dokkum}, P.~G., \& {Coppi}, P. 2008, \apj, 686, 1503
	
	\bibitem[{{Brammer} {et~al.}(2012){Brammer}, {van Dokkum}, {Franx},
		{Fumagalli}, {Patel}, {Rix}, {Skelton}, {Kriek}, {Nelson}, {Schmidt},
		{Bezanson}, {da Cunha}, {Erb}, {Fan}, {F{\"o}rster Schreiber}, {Illingworth},
		{Labb{\'e}}, {Leja}, {Lundgren}, {Magee}, {Marchesini}, {McCarthy},
		{Momcheva}, {Muzzin}, {Quadri}, {Steidel}, {Tal}, {Wake}, {Whitaker}, \&
		{Williams}}]{brammer2012}
	{Brammer}, G.~B., {van Dokkum}, P.~G., {Franx}, M., {et~al.} 2012, \apjs, 200,
	13
	
	\bibitem[{{Bruzual} \& {Charlot}(2003)}]{bruzual2003}
	{Bruzual}, G., \& {Charlot}, S. 2003, \mnras, 344, 1000
	
	\bibitem[{{Bussmann} {et~al.}(2008){Bussmann}, {Narayanan}, {Shirley},
		{Juneau}, {Wu}, {Solomon}, {Vanden Bout}, {Moustakas}, \&
		{Walker}}]{bussmann2008}
	{Bussmann}, R.~S., {Narayanan}, D., {Shirley}, Y.~L., {et~al.} 2008, \apjl,
	681, L73
	
	\bibitem[{{Calzetti} {et~al.}(2000){Calzetti}, {Armus}, {Bohlin}, {Kinney},
		{Koornneef}, \& {Storchi-Bergmann}}]{calzetti2000}
	{Calzetti}, D., {Armus}, L., {Bohlin}, R.~C., {et~al.} 2000, \apj, 533, 682
	
	\bibitem[{{Carilli} \& {Walter}(2013)}]{carilli2013}
	{Carilli}, C.~L., \& {Walter}, F. 2013, \araa, 51, 105
	
	\bibitem[{{Carilli} {et~al.}(2005){Carilli}, {Solomon}, {Vanden Bout},
		{Walter}, {Beelen}, {Cox}, {Bertoldi}, {Menten}, {Isaak}, {Chandler}, \&
		{Omont}}]{carilli2005}
	{Carilli}, C.~L., {Solomon}, P., {Vanden Bout}, P., {et~al.} 2005, \apj, 618,
	586
	
	\bibitem[{{Carleton} {et~al.}(2016){Carleton}, {Cooper}, {Bolatto}, {Bournaud},
		{Combes}, {Freundlich}, {Garcia-Burillo}, {Genzel}, {Neri}, {Tacconi},
		{Sandstrom}, {Weiner}, \& {Weiss}}]{carleton2016}
	{Carleton}, T., {Cooper}, M.~C., {Bolatto}, A.~D., {et~al.} 2016, ArXiv
	e-prints, arXiv:1611.04587
	
	\bibitem[{{Casey}(2012)}]{casey2012}
	{Casey}, C.~M. 2012, \mnras, 425, 3094
	
	\bibitem[{{Casey} {et~al.}(2014){Casey}, {Narayanan}, \& {Cooray}}]{casey2014}
	{Casey}, C.~M., {Narayanan}, D., \& {Cooray}, A. 2014, \physrep, 541, 45
	
	\bibitem[{{Charlot} \& {Fall}(2000)}]{charlot2000}
	{Charlot}, S., \& {Fall}, S.~M. 2000, \apj, 539, 718
	
	\bibitem[{{Costagliola} {et~al.}(2011){Costagliola}, {Aalto}, {Rodriguez},
		{Muller}, {Spoon}, {Mart{\'{\i}}n}, {Per{\'e}z-Torres}, {Alberdi},
		{Lindberg}, {Batejat}, {J{\"u}tte}, {van der Werf}, \&
		{Lahuis}}]{costagliola2011}
	{Costagliola}, F., {Aalto}, S., {Rodriguez}, M.~I., {et~al.} 2011, \aap, 528,
	A30
	
	\bibitem[{{da Cunha} {et~al.}(2008){da Cunha}, {Charlot}, \&
		{Elbaz}}]{dacunha2008}
	{da Cunha}, E., {Charlot}, S., \& {Elbaz}, D. 2008, \mnras, 388, 1595
	
	\bibitem[{{da Cunha} {et~al.}(2015){da Cunha}, {Walter}, {Smail}, {Swinbank},
		{Simpson}, {Decarli}, {Hodge}, {Weiss}, {van der Werf}, {Bertoldi},
		{Chapman}, {Cox}, {Danielson}, {Dannerbauer}, {Greve}, {Ivison}, {Karim}, \&
		{Thomson}}]{dacunha2015}
	{da Cunha}, E., {Walter}, F., {Smail}, I.~R., {et~al.} 2015, \apj, 806, 110
	
	\bibitem[{{Daddi} {et~al.}(2008){Daddi}, {Dannerbauer}, {Elbaz}, {Dickinson},
		{Morrison}, {Stern}, \& {Ravindranath}}]{daddi2008}
	{Daddi}, E., {Dannerbauer}, H., {Elbaz}, D., {et~al.} 2008, \apjl, 673, L21
	
	\bibitem[{{Daddi} {et~al.}(2007){Daddi}, {Dickinson}, {Morrison}, {Chary},
		{Cimatti}, {Elbaz}, {Frayer}, {Renzini}, {Pope}, {Alexander}, {Bauer},
		{Giavalisco}, {Huynh}, {Kurk}, \& {Mignoli}}]{daddi2007}
	{Daddi}, E., {Dickinson}, M., {Morrison}, G., {et~al.} 2007, \apj, 670, 156
	
	\bibitem[{{Daddi} {et~al.}(2010{\natexlab{a}}){Daddi}, {Elbaz}, {Walter},
		{Bournaud}, {Salmi}, {Carilli}, {Dannerbauer}, {Dickinson}, {Monaco}, \&
		{Riechers}}]{daddi2010b}
	{Daddi}, E., {Elbaz}, D., {Walter}, F., {et~al.} 2010{\natexlab{a}}, \apjl,
	714, L118
	
	\bibitem[{{Daddi} {et~al.}(2010{\natexlab{b}}){Daddi}, {Bournaud}, {Walter},
		{Dannerbauer}, {Carilli}, {Dickinson}, {Elbaz}, {Morrison}, {Riechers},
		{Onodera}, {Salmi}, {Krips}, \& {Stern}}]{daddi2010a}
	{Daddi}, E., {Bournaud}, F., {Walter}, F., {et~al.} 2010{\natexlab{b}}, \apj,
	713, 686
	
	\bibitem[{{Daddi} {et~al.}(2015){Daddi}, {Dannerbauer}, {Liu}, {Aravena},
		{Bournaud}, {Walter}, {Riechers}, {Magdis}, {Sargent}, {B{\'e}thermin},
		{Carilli}, {Cibinel}, {Dickinson}, {Elbaz}, {Gao}, {Gobat}, {Hodge}, \&
		{Krips}}]{daddi2015}
	{Daddi}, E., {Dannerbauer}, H., {Liu}, D., {et~al.} 2015, \aap, 577, A46
	
	\bibitem[{{Dale} {et~al.}(2014){Dale}, {Helou}, {Magdis}, {Armus},
		{D{\'{\i}}az-Santos}, \& {Shi}}]{dale2014}
	{Dale}, D.~A., {Helou}, G., {Magdis}, G.~E., {et~al.} 2014, \apj, 784, 83
	
	\bibitem[{{Danielson} {et~al.}(2011){Danielson}, {Swinbank}, {Smail}, {Cox},
		{Edge}, {Weiss}, {Harris}, {Baker}, {De Breuck}, {Geach}, {Ivison}, {Krips},
		{Lundgren}, {Longmore}, {Neri}, \& {Flaquer}}]{danielson2011}
	{Danielson}, A.~L.~R., {Swinbank}, A.~M., {Smail}, I., {et~al.} 2011, \mnras,
	410, 1687
	
	\bibitem[{{Dannerbauer} {et~al.}(2009){Dannerbauer}, {Daddi}, {Riechers},
		{Walter}, {Carilli}, {Dickinson}, {Elbaz}, \& {Morrison}}]{dannerbauer2009}
	{Dannerbauer}, H., {Daddi}, E., {Riechers}, D.~A., {et~al.} 2009, \apjl, 698,
	L178
	
	\bibitem[{{Davis} {et~al.}(2007){Davis}, {Guhathakurta}, {Konidaris}, {Newman},
		{Ashby}, {Biggs}, {Barmby}, {Bundy}, {Chapman}, {Coil}, {Conselice},
		{Cooper}, {Croton}, {Eisenhardt}, {Ellis}, {Faber}, {Fang}, {Fazio},
		{Georgakakis}, {Gerke}, {Goss}, {Gwyn}, {Harker}, {Hopkins}, {Huang},
		{Ivison}, {Kassin}, {Kirby}, {Koekemoer}, {Koo}, {Laird}, {Le Floc'h}, {Lin},
		{Lotz}, {Marshall}, {Martin}, {Metevier}, {Moustakas}, {Nandra}, {Noeske},
		{Papovich}, {Phillips}, {Rich}, {Rieke}, {Rigopoulou}, {Salim},
		{Schiminovich}, {Simard}, {Smail}, {Small}, {Weiner}, {Willmer}, {Willner},
		{Wilson}, {Wright}, \& {Yan}}]{davis2007}
	{Davis}, M., {Guhathakurta}, P., {Konidaris}, N.~P., {et~al.} 2007, \apjl, 660,
	L1
	
	\bibitem[{{Dessauges-Zavadsky} {et~al.}(2015){Dessauges-Zavadsky}, {Zamojski},
		{Schaerer}, {Combes}, {Egami}, {Swinbank}, {Richard}, {Sklias}, {Rawle},
		{Rex}, {Kneib}, {Boone}, \& {Blain}}]{dz2015}
	{Dessauges-Zavadsky}, M., {Zamojski}, M., {Schaerer}, D., {et~al.} 2015, \aap,
	577, A50
	
	\bibitem[{{Downes} \& {Solomon}(1998)}]{downes1998}
	{Downes}, D., \& {Solomon}, P.~M. 1998, \apj, 507, 615
	
	\bibitem[{{Draine} \& {Li}(2007)}]{draine2007}
	{Draine}, B.~T., \& {Li}, A. 2007, \apj, 657, 810
	
	\bibitem[{{Elbaz} {et~al.}(2007){Elbaz}, {Daddi}, {Le Borgne}, {Dickinson},
		{Alexander}, {Chary}, {Starck}, {Brandt}, {Kitzbichler}, {MacDonald},
		{Nonino}, {Popesso}, {Stern}, \& {Vanzella}}]{elbaz2007}
	{Elbaz}, D., {Daddi}, E., {Le Borgne}, D., {et~al.} 2007, \aap, 468, 33
	
	\bibitem[{{Engel} {et~al.}(2010){Engel}, {Tacconi}, {Davies}, {Neri}, {Smail},
		{Chapman}, {Genzel}, {Cox}, {Greve}, {Ivison}, {Blain}, {Bertoldi}, \&
		{Omont}}]{engel2010}
	{Engel}, H., {Tacconi}, L.~J., {Davies}, R.~I., {et~al.} 2010, \apj, 724, 233
	
	\bibitem[{{Evans} {et~al.}(2006){Evans}, {Solomon}, {Tacconi}, {Vavilkin}, \&
		{Downes}}]{evans2006}
	{Evans}, A.~S., {Solomon}, P.~M., {Tacconi}, L.~J., {Vavilkin}, T., \&
	{Downes}, D. 2006, \aj, 132, 2398
	
	\bibitem[{{Freundlich} {et~al.}(2013){Freundlich}, {Combes}, {Tacconi},
		{Cooper}, {Genzel}, {Neri}, {Bolatto}, {Bournaud}, {Burkert}, {Cox}, {Davis},
		{F{\"o}rster Schreiber}, {Garcia-Burillo}, {Gracia-Carpio}, {Lutz}, {Naab},
		{Newman}, {Sternberg}, \& {Weiner}}]{freundlich2013}
	{Freundlich}, J., {Combes}, F., {Tacconi}, L.~J., {et~al.} 2013, \aap, 553,
	A130
	
	\bibitem[{{Fritz} {et~al.}(2006){Fritz}, {Franceschini}, \&
		{Hatziminaoglou}}]{fritz2006}
	{Fritz}, J., {Franceschini}, A., \& {Hatziminaoglou}, E. 2006, \mnras, 366, 767
	
	\bibitem[{{Gao} {et~al.}(2007){Gao}, {Carilli}, {Solomon}, \& {Vanden
			Bout}}]{gao2007}
	{Gao}, Y., {Carilli}, C.~L., {Solomon}, P.~M., \& {Vanden Bout}, P.~A. 2007,
	\apjl, 660, L93
	
	\bibitem[{{Gao} {et~al.}(2001){Gao}, {Lo}, {Lee}, \& {Lee}}]{gao2001}
	{Gao}, Y., {Lo}, K.~Y., {Lee}, S.-W., \& {Lee}, T.-H. 2001, \apj, 548, 172
	
	\bibitem[{{Gao} \& {Solomon}(2004{\natexlab{a}})}]{gao2004a}
	{Gao}, Y., \& {Solomon}, P.~M. 2004{\natexlab{a}}, \apjs, 152, 63
	
	\bibitem[{{Gao} \& {Solomon}(2004{\natexlab{b}})}]{gao2004b}
	---. 2004{\natexlab{b}}, \apj, 606, 271
	
	\bibitem[{{Garc{\'{\i}}a-Burillo} {et~al.}(2012){Garc{\'{\i}}a-Burillo},
		{Usero}, {Alonso-Herrero}, {Graci{\'a}-Carpio}, {Pereira-Santaella},
		{Colina}, {Planesas}, \& {Arribas}}]{garcia2012}
	{Garc{\'{\i}}a-Burillo}, S., {Usero}, A., {Alonso-Herrero}, A., {et~al.} 2012,
	\aap, 539, A8
	
	\bibitem[{{Geach} \& {Papadopoulos}(2012)}]{geach2012}
	{Geach}, J.~E., \& {Papadopoulos}, P.~P. 2012, \apj, 757, 156
	
	\bibitem[{{Genzel} {et~al.}(2010){Genzel}, {Tacconi}, {Gracia-Carpio},
		{Sternberg}, {Cooper}, {Shapiro}, {Bolatto}, {Bouch{\'e}}, {Bournaud},
		{Burkert}, {Combes}, {Comerford}, {Cox}, {Davis}, {Schreiber},
		{Garcia-Burillo}, {Lutz}, {Naab}, {Neri}, {Omont}, {Shapley}, \&
		{Weiner}}]{genzel2010}
	{Genzel}, R., {Tacconi}, L.~J., {Gracia-Carpio}, J., {et~al.} 2010, \mnras,
	407, 2091
	
	\bibitem[{{Genzel} {et~al.}(2015){Genzel}, {Tacconi}, {Lutz}, {Saintonge},
		{Berta}, {Magnelli}, {Combes}, {Garc{\'{\i}}a-Burillo}, {Neri}, {Bolatto},
		{Contini}, {Lilly}, {Boissier}, {Boone}, {Bouch{\'e}}, {Bournaud}, {Burkert},
		{Carollo}, {Colina}, {Cooper}, {Cox}, {Feruglio}, {F{\"o}rster Schreiber},
		{Freundlich}, {Gracia-Carpio}, {Juneau}, {Kovac}, {Lippa}, {Naab}, {Salome},
		{Renzini}, {Sternberg}, {Walter}, {Weiner}, {Weiss}, \& {Wuyts}}]{genzel2015}
	{Genzel}, R., {Tacconi}, L.~J., {Lutz}, D., {et~al.} 2015, \apj, 800, 20
	
	\bibitem[{{Goldsmith} {et~al.}(1981){Goldsmith}, {Langer}, {Ellder},
		{Kollberg}, \& {Irvine}}]{goldsmith1981}
	{Goldsmith}, P.~F., {Langer}, W.~D., {Ellder}, J., {Kollberg}, E., \& {Irvine},
	W. 1981, \apj, 249, 524
	
	\bibitem[{{Graci{\'a}-Carpio} {et~al.}(2006){Graci{\'a}-Carpio},
		{Garc{\'{\i}}a-Burillo}, {Planesas}, \& {Colina}}]{gracia2006}
	{Graci{\'a}-Carpio}, J., {Garc{\'{\i}}a-Burillo}, S., {Planesas}, P., \&
	{Colina}, L. 2006, \apjl, 640, L135
	
	\bibitem[{{Graci{\'a}-Carpio} {et~al.}(2008){Graci{\'a}-Carpio},
		{Garc{\'{\i}}a-Burillo}, {Planesas}, {Fuente}, \& {Usero}}]{gracia2008}
	{Graci{\'a}-Carpio}, J., {Garc{\'{\i}}a-Burillo}, S., {Planesas}, P., {Fuente},
	A., \& {Usero}, A. 2008, \aap, 479, 703
	
	\bibitem[{{Graninger} {et~al.}(2014){Graninger}, {Herbst}, {{\"O}berg}, \&
		{Vasyunin}}]{graninger2014}
	{Graninger}, D.~M., {Herbst}, E., {{\"O}berg}, K.~I., \& {Vasyunin}, A.~I.
	2014, \apj, 787, 74
	
	\bibitem[{{Greve} {et~al.}(2006){Greve}, {Hainline}, {Blain}, {Smail},
		{Ivison}, \& {Papadopoulos}}]{greve2006}
	{Greve}, T.~R., {Hainline}, L.~J., {Blain}, A.~W., {et~al.} 2006, \aj, 132,
	1938
	
	\bibitem[{{Greve} {et~al.}(2005){Greve}, {Bertoldi}, {Smail}, {Neri},
		{Chapman}, {Blain}, {Ivison}, {Genzel}, {Omont}, {Cox}, {Tacconi}, \&
		{Kneib}}]{greve2005}
	{Greve}, T.~R., {Bertoldi}, F., {Smail}, I., {et~al.} 2005, \mnras, 359, 1165
	
	\bibitem[{{Groves} {et~al.}(2015){Groves}, {Schinnerer}, {Leroy}, {Galametz},
		{Walter}, {Bolatto}, {Hunt}, {Dale}, {Calzetti}, {Croxall}, \&
		{Kennicutt}}]{groves2015}
	{Groves}, B.~A., {Schinnerer}, E., {Leroy}, A., {et~al.} 2015, \apj, 799, 96
	
	\bibitem[{{Hodge} {et~al.}(2012){Hodge}, {Carilli}, {Walter}, {de Blok},
		{Riechers}, {Daddi}, \& {Lentati}}]{hodge2012}
	{Hodge}, J.~A., {Carilli}, C.~L., {Walter}, F., {et~al.} 2012, \apj, 760, 11
	
	\bibitem[{{Huang} {et~al.}(2009){Huang}, {Faber}, {Daddi}, {Laird}, {Lai},
		{Omont}, {Wu}, {Younger}, {Bundy}, {Cattaneo}, {Chapman}, {Conselice},
		{Dickinson}, {Egami}, {Fazio}, {Im}, {Koo}, {Le Floc'h}, {Papovich},
		{Rigopoulou}, {Smail}, {Song}, {Van de Werf}, {Webb}, {Willmer}, {Willner},
		\& {Yan}}]{huang2009}
	{Huang}, J.-S., {Faber}, S.~M., {Daddi}, E., {et~al.} 2009, \apj, 700, 183
	
	\bibitem[{{Huettemeister} {et~al.}(1995){Huettemeister}, {Henkel},
		{Mauersberger}, {Brouillet}, {Wiklind}, \& {Millar}}]{huett1995}
	{Huettemeister}, S., {Henkel}, C., {Mauersberger}, R., {et~al.} 1995, \aap,
	295, 571
	
	\bibitem[{{Imanishi} \& {Nakanishi}(2014)}]{imanishi2014}
	{Imanishi}, M., \& {Nakanishi}, K. 2014, \aj, 148, 9
	
	\bibitem[{{Imanishi} {et~al.}(2007){Imanishi}, {Nakanishi}, {Tamura}, {Oi}, \&
		{Kohno}}]{imanishi2007}
	{Imanishi}, M., {Nakanishi}, K., {Tamura}, Y., {Oi}, N., \& {Kohno}, K. 2007,
	\aj, 134, 2366
	
	\bibitem[{{Isaak} {et~al.}(2004){Isaak}, {Chandler}, \& {Carilli}}]{isaak2004}
	{Isaak}, K.~G., {Chandler}, C.~J., \& {Carilli}, C.~L. 2004, \mnras, 348, 1035
	
	\bibitem[{{Ivison} {et~al.}(2011){Ivison}, {Papadopoulos}, {Smail}, {Greve},
		{Thomson}, {Xilouris}, \& {Chapman}}]{ivison2011}
	{Ivison}, R.~J., {Papadopoulos}, P.~P., {Smail}, I., {et~al.} 2011, \mnras,
	412, 1913
	
	\bibitem[{{Izumi} {et~al.}(2015){Izumi}, {Kohno}, {Aalto}, {Doi}, {Espada},
		{Fathi}, {Harada}, {Hatsukade}, {Hattori}, {Hsieh}, {Ikarashi}, {Imanishi},
		{Iono}, {Ishizuki}, {Krips}, {Mart{\'{\i}}n}, {Matsushita}, {Meier}, {Nagai},
		{Nakai}, {Nakajima}, {Nakanishi}, {Nomura}, {Regan}, {Schinnerer}, {Sheth},
		{Takano}, {Tamura}, {Terashima}, {Tosaki}, {Turner}, {Umehata}, \&
		{Wiklind}}]{izumi2015}
	{Izumi}, T., {Kohno}, K., {Aalto}, S., {et~al.} 2015, \apj, 811, 39
	
	\bibitem[{{Izumi} {et~al.}(2016){Izumi}, {Kohno}, {Aalto}, {Espada}, {Fathi},
		{Harada}, {Hatsukade}, {Hsieh}, {Imanishi}, {Krips}, {Mart{\'{\i}}n},
		{Matsushita}, {Meier}, {Nakai}, {Nakanishi}, {Schinnerer}, {Sheth},
		{Terashima}, \& {Turner}}]{izumi2016}
	---. 2016, \apj, 818, 42
	
	\bibitem[{{Juneau} {et~al.}(2009){Juneau}, {Narayanan}, {Moustakas}, {Shirley},
		{Bussmann}, {Kennicutt}, \& {Vanden Bout}}]{juneau2009}
	{Juneau}, S., {Narayanan}, D.~T., {Moustakas}, J., {et~al.} 2009, \apj, 707,
	1217
	
	\bibitem[{{Karim} {et~al.}(2011){Karim}, {Schinnerer},
		{Mart{\'{\i}}nez-Sansigre}, {Sargent}, {van der Wel}, {Rix}, {Ilbert},
		{Smol{\v c}i{\'c}}, {Carilli}, {Pannella}, {Koekemoer}, {Bell}, \&
		{Salvato}}]{karim2011}
	{Karim}, A., {Schinnerer}, E., {Mart{\'{\i}}nez-Sansigre}, A., {et~al.} 2011,
	\apj, 730, 61
	
	\bibitem[{{Kennicutt}(1998)}]{kennicutt1998}
	{Kennicutt}, Jr., R.~C. 1998, \apj, 498, 541
	
	\bibitem[{{Kewley} {et~al.}(2002){Kewley}, {Geller}, {Jansen}, \&
		{Dopita}}]{kewley2002}
	{Kewley}, L.~J., {Geller}, M.~J., {Jansen}, R.~A., \& {Dopita}, M.~A. 2002,
	\aj, 124, 3135
	
	\bibitem[{{Kohno} {et~al.}(2001){Kohno}, {Matsushita}, {Vila-Vilar{\'o}},
		{Okumura}, {Shibatsuka}, {Okiura}, {Ishizuki}, \& {Kawabe}}]{kohno2001}
	{Kohno}, K., {Matsushita}, S., {Vila-Vilar{\'o}}, B., {et~al.} 2001, in
	Astronomical Society of the Pacific Conference Series, Vol. 249, The Central
	Kiloparsec of Starbursts and AGN: The La Palma Connection, ed. J.~H.
	{Knapen}, J.~E. {Beckman}, I.~{Shlosman}, \& T.~J. {Mahoney}, 672
	
	\bibitem[{{Krips} {et~al.}(2010){Krips}, {Crocker}, {Bureau}, {Combes}, \&
		{Young}}]{krips2010}
	{Krips}, M., {Crocker}, A.~F., {Bureau}, M., {Combes}, F., \& {Young}, L.~M.
	2010, \mnras, 407, 2261
	
	\bibitem[{{Krips} {et~al.}(2008){Krips}, {Neri}, {Garc{\'{\i}}a-Burillo},
		{Mart{\'{\i}}n}, {Combes}, {Graci{\'a}-Carpio}, \& {Eckart}}]{krips2008}
	{Krips}, M., {Neri}, R., {Garc{\'{\i}}a-Burillo}, S., {et~al.} 2008, \apj, 677,
	262
	
	\bibitem[{{Krumholz} \& {McKee}(2005)}]{krumholz2005}
	{Krumholz}, M.~R., \& {McKee}, C.~F. 2005, \apj, 630, 250
	
	\bibitem[{{Krumholz} \& {Thompson}(2007)}]{krumholz2007b}
	{Krumholz}, M.~R., \& {Thompson}, T.~A. 2007, \apj, 669, 289
	
	\bibitem[{{Lada} {et~al.}(2012){Lada}, {Forbrich}, {Lombardi}, \&
		{Alves}}]{lada2012}
	{Lada}, C.~J., {Forbrich}, J., {Lombardi}, M., \& {Alves}, J.~F. 2012, \apj,
	745, 190
	
	\bibitem[{{Laird} {et~al.}(2009){Laird}, {Nandra}, {Georgakakis}, {Aird},
		{Barmby}, {Conselice}, {Coil}, {Davis}, {Faber}, {Fazio}, {Guhathakurta},
		{Koo}, {Sarajedini}, \& {Willmer}}]{laird2009}
	{Laird}, E.~S., {Nandra}, K., {Georgakakis}, A., {et~al.} 2009, \apjs, 180, 102
	
	\bibitem[{{Lee} {et~al.}(2015){Lee}, {Sanders}, {Casey}, {Toft}, {Scoville},
		{Hung}, {Le Floc'h}, {Ilbert}, {Zahid}, {Aussel}, {Capak}, {Kartaltepe},
		{Kewley}, {Li}, {Schawinski}, {Sheth}, \& {Xiao}}]{lee2015}
	{Lee}, N., {Sanders}, D.~B., {Casey}, C.~M., {et~al.} 2015, \apj, 801, 80
	
	\bibitem[{{Lepp} \& {Dalgarno}(1996)}]{lepp1996}
	{Lepp}, S., \& {Dalgarno}, A. 1996, \aap, 306, L21
	
	\bibitem[{{Leroy} {et~al.}(2011){Leroy}, {Bolatto}, {Gordon}, {Sandstrom},
		{Gratier}, {Rosolowsky}, {Engelbracht}, {Mizuno}, {Corbelli}, {Fukui}, \&
		{Kawamura}}]{leroy2011}
	{Leroy}, A.~K., {Bolatto}, A., {Gordon}, K., {et~al.} 2011, \apj, 737, 12
	
	\bibitem[{{Lilly} {et~al.}(2013){Lilly}, {Carollo}, {Pipino}, {Renzini}, \&
		{Peng}}]{lilly2013}
	{Lilly}, S.~J., {Carollo}, C.~M., {Pipino}, A., {Renzini}, A., \& {Peng}, Y.
	2013, \apj, 772, 119
	
	\bibitem[{{Lindberg} {et~al.}(2016){Lindberg}, {Aalto}, {Muller},
		{Mart{\'{\i}}-Vidal}, {Falstad}, {Costagliola}, {Henkel}, {van der Werf},
		{Garc{\'{\i}}a-Burillo}, \& {Gonz{\'a}lez-Alfonso}}]{lindberg2016}
	{Lindberg}, J.~E., {Aalto}, S., {Muller}, S., {et~al.} 2016, \aap, 587, A15
	
	\bibitem[{{Loenen} {et~al.}(2007){Loenen}, {Baan}, \& {Spaans}}]{loenen2007}
	{Loenen}, A.~F., {Baan}, W.~A., \& {Spaans}, M. 2007, in IAU Symposium, Vol.
	242, Astrophysical Masers and their Environments, ed. J.~M. {Chapman} \&
	W.~A. {Baan}, 462--466
	
	\bibitem[{{Magdis} {et~al.}(2011){Magdis}, {Daddi}, {Elbaz}, {Sargent},
		{Dickinson}, {Dannerbauer}, {Aussel}, {Walter}, {Hwang}, {Charmandaris},
		{Hodge}, {Riechers}, {Rigopoulou}, {Carilli}, {Pannella}, {Mullaney},
		{Leiton}, \& {Scott}}]{magdis2011}
	{Magdis}, G.~E., {Daddi}, E., {Elbaz}, D., {et~al.} 2011, \apjl, 740, L15
	
	\bibitem[{{Maraston}(2005)}]{maraston2005}
	{Maraston}, C. 2005, \mnras, 362, 799
	
	\bibitem[{{Mart{\'{\i}}n} {et~al.}(2015){Mart{\'{\i}}n}, {Kohno}, {Izumi},
		{Krips}, {Meier}, {Aladro}, {Matsushita}, {Takano}, {Turner}, {Espada},
		{Nakajima}, {Terashima}, {Fathi}, {Hsieh}, {Imanishi}, {Lundgren}, {Nakai},
		{Schinnerer}, {Sheth}, \& {Wiklind}}]{martin2015}
	{Mart{\'{\i}}n}, S., {Kohno}, K., {Izumi}, T., {et~al.} 2015, \aap, 573, A116
	
	\bibitem[{{Meier} \& {Turner}(2005)}]{meier2005}
	{Meier}, D.~S., \& {Turner}, J.~L. 2005, \apj, 618, 259
	
	\bibitem[{{Meijerink} {et~al.}(2007){Meijerink}, {Spaans}, \&
		{Israel}}]{meijerink2007}
	{Meijerink}, R., {Spaans}, M., \& {Israel}, F.~P. 2007, \aap, 461, 793
	
	\bibitem[{{Nandra} {et~al.}(2015){Nandra}, {Laird}, {Aird}, {Salvato},
		{Georgakakis}, {Barro}, {Perez-Gonzalez}, {Barmby}, {Chary}, {Coil},
		{Cooper}, {Davis}, {Dickinson}, {Faber}, {Fazio}, {Guhathakurta}, {Gwyn},
		{Hsu}, {Huang}, {Ivison}, {Koo}, {Newman}, {Rangel}, {Yamada}, \&
		{Willmer}}]{nandra2015}
	{Nandra}, K., {Laird}, E.~S., {Aird}, J.~A., {et~al.} 2015, \apjs, 220, 10
	
	\bibitem[{{Noeske} {et~al.}(2007){Noeske}, {Faber}, {Weiner}, {Koo}, {Primack},
		{Dekel}, {Papovich}, {Conselice}, {Le Floc'h}, {Rieke}, {Coil}, {Lotz},
		{Somerville}, \& {Bundy}}]{noeske2007}
	{Noeske}, K.~G., {Faber}, S.~M., {Weiner}, B.~J., {et~al.} 2007, \apjl, 660,
	L47
	
	\bibitem[{{Noll} {et~al.}(2009){Noll}, {Burgarella}, {Giovannoli}, {Buat},
		{Marcillac}, \& {Mu{\~n}oz-Mateos}}]{noll2009}
	{Noll}, S., {Burgarella}, D., {Giovannoli}, E., {et~al.} 2009, \aap, 507, 1793
	
	\bibitem[{Papadopoulos(2007)}]{papadopoulos2007}
	Papadopoulos, P.~P. 2007, The Astrophysical Journal, 656, 792
	
	\bibitem[{{Papadopoulos} {et~al.}(2012){Papadopoulos}, {van der Werf},
		{Xilouris}, {Isaak}, {Gao}, \& {M{\"u}hle}}]{papa2012}
	{Papadopoulos}, P.~P., {van der Werf}, P.~P., {Xilouris}, E.~M., {et~al.} 2012,
	\mnras, 426, 2601
	
	\bibitem[{{Pettini} {et~al.}(2001){Pettini}, {Shapley}, {Steidel}, {Cuby},
		{Dickinson}, {Moorwood}, {Adelberger}, \& {Giavalisco}}]{pettini2001}
	{Pettini}, M., {Shapley}, A.~E., {Steidel}, C.~C., {et~al.} 2001, \apj, 554,
	981
	
	\bibitem[{{Privon} {et~al.}(2015){Privon}, {Herrero-Illana}, {Evans},
		{Iwasawa}, {Perez-Torres}, {Armus}, {D{\'{\i}}az-Santos}, {Murphy},
		{Stierwalt}, {Aalto}, {Mazzarella}, {Barcos-Mu{\~n}oz}, {Borish}, {Inami},
		{Kim}, {Treister}, {Surace}, {Lord}, {Conway}, {Frayer}, \&
		{Alberdi}}]{privon2015}
	{Privon}, G.~C., {Herrero-Illana}, R., {Evans}, A.~S., {et~al.} 2015, \apj,
	814, 39
	
	\bibitem[{{R{\'e}my-Ruyer} {et~al.}(2014){R{\'e}my-Ruyer}, {Madden},
		{Galliano}, {Galametz}, {Takeuchi}, {Asano}, {Zhukovska}, {Lebouteiller},
		{Cormier}, {Jones}, {Bocchio}, {Baes}, {Bendo}, {Boquien}, {Boselli},
		{DeLooze}, {Doublier-Pritchard}, {Hughes}, {Karczewski}, \&
		{Spinoglio}}]{remy2014}
	{R{\'e}my-Ruyer}, A., {Madden}, S.~C., {Galliano}, F., {et~al.} 2014, \aap,
	563, A31
	
	\bibitem[{{Riechers} {et~al.}(2010{\natexlab{a}}){Riechers}, {Carilli},
		{Walter}, \& {Momjian}}]{riechers2010}
	{Riechers}, D.~A., {Carilli}, C.~L., {Walter}, F., \& {Momjian}, E.
	2010{\natexlab{a}}, \apjl, 724, L153
	
	\bibitem[{{Riechers} {et~al.}(2011{\natexlab{a}}){Riechers}, {Hodge}, {Walter},
		{Carilli}, \& {Bertoldi}}]{riechers2011b}
	{Riechers}, D.~A., {Hodge}, J., {Walter}, F., {Carilli}, C.~L., \& {Bertoldi},
	F. 2011{\natexlab{a}}, \apjl, 739, L31
	
	\bibitem[{{Riechers} {et~al.}(2007){Riechers}, {Walter}, {Carilli}, \&
		{Bertoldi}}]{riechers2007}
	{Riechers}, D.~A., {Walter}, F., {Carilli}, C.~L., \& {Bertoldi}, F. 2007,
	\apjl, 671, L13
	
	\bibitem[{{Riechers} {et~al.}(2009{\natexlab{a}}){Riechers}, {Walter},
		{Carilli}, \& {Lewis}}]{riechers2009}
	{Riechers}, D.~A., {Walter}, F., {Carilli}, C.~L., \& {Lewis}, G.~F.
	2009{\natexlab{a}}, \apj, 690, 463
	
	\bibitem[{{Riechers} {et~al.}(2006{\natexlab{a}}){Riechers}, {Walter},
		{Carilli}, {Weiss}, {Bertoldi}, {Menten}, {Knudsen}, \&
		{Cox}}]{riechers2006a}
	{Riechers}, D.~A., {Walter}, F., {Carilli}, C.~L., {et~al.} 2006{\natexlab{a}},
	\apjl, 645, L13
	
	\bibitem[{{Riechers} {et~al.}(2010{\natexlab{b}}){Riechers}, {Wei{\ss}},
		{Walter}, \& {Wagg}}]{riechers2010b}
	{Riechers}, D.~A., {Wei{\ss}}, A., {Walter}, F., \& {Wagg}, J.
	2010{\natexlab{b}}, \apj, 725, 1032
	
	\bibitem[{{Riechers} {et~al.}(2006{\natexlab{b}}){Riechers}, {Walter},
		{Carilli}, {Knudsen}, {Lo}, {Benford}, {Staguhn}, {Hunter}, {Bertoldi},
		{Henkel}, {Menten}, {Weiss}, {Yun}, \& {Scoville}}]{riechers2006b}
	{Riechers}, D.~A., {Walter}, F., {Carilli}, C.~L., {et~al.} 2006{\natexlab{b}},
	\apj, 650, 604
	
	\bibitem[{{Riechers} {et~al.}(2009{\natexlab{b}}){Riechers}, {Walter},
		{Bertoldi}, {Carilli}, {Aravena}, {Neri}, {Cox}, {Wei{\ss}}, \&
		{Menten}}]{riechers2009a}
	{Riechers}, D.~A., {Walter}, F., {Bertoldi}, F., {et~al.} 2009{\natexlab{b}},
	\apj, 703, 1338
	
	\bibitem[{{Riechers} {et~al.}(2011{\natexlab{b}}){Riechers}, {Carilli},
		{Maddalena}, {Hodge}, {Harris}, {Baker}, {Walter}, {Wagg}, {Vanden Bout},
		{Wei{\ss}}, \& {Sharon}}]{riechers2011}
	{Riechers}, D.~A., {Carilli}, C.~L., {Maddalena}, R.~J., {et~al.}
	2011{\natexlab{b}}, \apjl, 739, L32
	
	\bibitem[{{Riechers} {et~al.}(2013){Riechers}, {Bradford}, {Clements},
		{Dowell}, {P{\'e}rez-Fournon}, {Ivison}, {Bridge}, {Conley}, {Fu}, {Vieira},
		{Wardlow}, {Calanog}, {Cooray}, {Hurley}, {Neri}, {Kamenetzky}, {Aguirre},
		{Altieri}, {Arumugam}, {Benford}, {B{\'e}thermin}, {Bock}, {Burgarella},
		{Cabrera-Lavers}, {Chapman}, {Cox}, {Dunlop}, {Earle}, {Farrah}, {Ferrero},
		{Franceschini}, {Gavazzi}, {Glenn}, {Solares}, {Gurwell}, {Halpern},
		{Hatziminaoglou}, {Hyde}, {Ibar}, {Kov{\'a}cs}, {Krips}, {Lupu}, {Maloney},
		{Martinez-Navajas}, {Matsuhara}, {Murphy}, {Naylor}, {Nguyen}, {Oliver},
		{Omont}, {Page}, {Petitpas}, {Rangwala}, {Roseboom}, {Scott}, {Smith},
		{Staguhn}, {Streblyanska}, {Thomson}, {Valtchanov}, {Viero}, {Wang},
		{Zemcov}, \& {Zmuidzinas}}]{riechers2013}
	{Riechers}, D.~A., {Bradford}, C.~M., {Clements}, D.~L., {et~al.} 2013, \nat,
	496, 329
	
	\bibitem[{{Riechers} {et~al.}(2014){Riechers}, {Carilli}, {Capak}, {Scoville},
		{Smol{\v c}i{\'c}}, {Schinnerer}, {Yun}, {Cox}, {Bertoldi}, {Karim}, \&
		{Yan}}]{riechers2014}
	{Riechers}, D.~A., {Carilli}, C.~L., {Capak}, P.~L., {et~al.} 2014, \apj, 796,
	84
	
	\bibitem[{{Rodighiero} {et~al.}(2011){Rodighiero}, {Daddi}, {Baronchelli},
		{Cimatti}, {Renzini}, {Aussel}, {Popesso}, {Lutz}, {Andreani}, {Berta},
		{Cava}, {Elbaz}, {Feltre}, {Fontana}, {F{\"o}rster Schreiber},
		{Franceschini}, {Genzel}, {Grazian}, {Gruppioni}, {Ilbert}, {Le Floch},
		{Magdis}, {Magliocchetti}, {Magnelli}, {Maiolino}, {McCracken}, {Nordon},
		{Poglitsch}, {Santini}, {Pozzi}, {Riguccini}, {Tacconi}, {Wuyts}, \&
		{Zamorani}}]{rodighiero2011}
	{Rodighiero}, G., {Daddi}, E., {Baronchelli}, I., {et~al.} 2011, \apjl, 739,
	L40
	
	\bibitem[{{Sanders} \& {Mirabel}(1996)}]{sanders1996}
	{Sanders}, D.~B., \& {Mirabel}, I.~F. 1996, \araa, 34, 749
	
	\bibitem[{{Sandstrom} {et~al.}(2013){Sandstrom}, {Leroy}, {Walter}, {Bolatto},
		{Croxall}, {Draine}, {Wilson}, {Wolfire}, {Calzetti}, {Kennicutt}, {Aniano},
		{Donovan Meyer}, {Usero}, {Bigiel}, {Brinks}, {de Blok}, {Crocker}, {Dale},
		{Engelbracht}, {Galametz}, {Groves}, {Hunt}, {Koda}, {Kreckel}, {Linz},
		{Meidt}, {Pellegrini}, {Rix}, {Roussel}, {Schinnerer}, {Schruba}, {Schuster},
		{Skibba}, {van der Laan}, {Appleton}, {Armus}, {Brandl}, {Gordon}, {Hinz},
		{Krause}, {Montiel}, {Sauvage}, {Schmiedeke}, {Smith}, \&
		{Vigroux}}]{sandstrom2013}
	{Sandstrom}, K.~M., {Leroy}, A.~K., {Walter}, F., {et~al.} 2013, \apj, 777, 5
	
	\bibitem[{{Santini} {et~al.}(2010){Santini}, {Maiolino}, {Magnelli}, {Silva},
		{Grazian}, {Altieri}, {Andreani}, {Aussel}, {Berta}, {Bongiovanni},
		{Brisbin}, {Calura}, {Cava}, {Cepa}, {Cimatti}, {Daddi}, {Dannerbauer},
		{Dominguez-Sanchez}, {Elbaz}, {Fontana}, {F{\"o}rster Schreiber}, {Genzel},
		{Granato}, {Gruppioni}, {Lutz}, {Magdis}, {Magliocchetti}, {Matteucci},
		{Nordon}, {P{\'e}rez Garcia}, {Poglitsch}, {Popesso}, {Pozzi}, {Riguccini},
		{Rodighiero}, {Saintonge}, {Sanchez-Portal}, {Shao}, {Sturm}, {Tacconi}, \&
		{Valtchanov}}]{santini2010}
	{Santini}, P., {Maiolino}, R., {Magnelli}, B., {et~al.} 2010, \aap, 518, L154
	
	\bibitem[{{Schilke} {et~al.}(1992){Schilke}, {Walmsley}, {Pineau Des Forets},
		{Roueff}, {Flower}, \& {Guilloteau}}]{schilke1992}
	{Schilke}, P., {Walmsley}, C.~M., {Pineau Des Forets}, G., {et~al.} 1992, \aap,
	256, 595
	
	\bibitem[{{Schmidt}(1959)}]{schmidt1959}
	{Schmidt}, M. 1959, \apj, 129, 243
	
	\bibitem[{{Scoville} {et~al.}(2015){Scoville}, {Sheth}, {Aussel}, {Vanden
			Bout}, {Capak}, {Bongiorno}, {Casey}, {Murchikova}, {Koda}, {Pope}, {Toft},
		{Ivison}, {Sanders}, {Manohar}, \& {Lee}}]{scoville2015}
	{Scoville}, N., {Sheth}, K., {Aussel}, H., {et~al.} 2015, ArXiv e-prints,
	arXiv:1505.02159
	
	\bibitem[{{Seko} {et~al.}(2016){Seko}, {Ohta}, {Yabe}, {Hatsukade}, {Akiyama},
		{Iwamuro}, {Tamura}, \& {Dalton}}]{seko2016}
	{Seko}, A., {Ohta}, K., {Yabe}, K., {et~al.} 2016, \apj, 819, 82
	
	\bibitem[{{Serra} {et~al.}(2011){Serra}, {Amblard}, {Temi}, {Burgarella},
		{Giovannoli}, {Buat}, {Noll}, \& {Im}}]{serra2011}
	{Serra}, P., {Amblard}, A., {Temi}, P., {et~al.} 2011, \apj, 740, 22
	
	\bibitem[{{Sharon} {et~al.}(2016){Sharon}, {Riechers}, {Hodge}, {Carilli},
		{Walter}, {Weiss}, {Knudsen}, \& {Wagg}}]{sharon2016}
	{Sharon}, C.~E., {Riechers}, D.~A., {Hodge}, J., {et~al.} 2016, ArXiv e-prints,
	arXiv:1606.02309
	
	\bibitem[{{Silverman} {et~al.}(2015){Silverman}, {Daddi}, {Rodighiero},
		{Rujopakarn}, {Sargent}, {Renzini}, {Liu}, {Feruglio}, {Kashino}, {Sanders},
		{Kartaltepe}, {Nagao}, {Arimoto}, {Berta}, {B{\'e}thermin}, {Koekemoer},
		{Lutz}, {Magdis}, {Mancini}, {Onodera}, \& {Zamorani}}]{silverman2015}
	{Silverman}, J.~D., {Daddi}, E., {Rodighiero}, G., {et~al.} 2015, \apjl, 812,
	L23
	
	\bibitem[{{Skelton} {et~al.}(2014){Skelton}, {Whitaker}, {Momcheva}, {Brammer},
		{van Dokkum}, {Labb{\'e}}, {Franx}, {van der Wel}, {Bezanson}, {Da Cunha},
		{Fumagalli}, {F{\"o}rster Schreiber}, {Kriek}, {Leja}, {Lundgren}, {Magee},
		{Marchesini}, {Maseda}, {Nelson}, {Oesch}, {Pacifici}, {Patel}, {Price},
		{Rix}, {Tal}, {Wake}, \& {Wuyts}}]{skelton2014}
	{Skelton}, R.~E., {Whitaker}, K.~E., {Momcheva}, I.~G., {et~al.} 2014, \apjs,
	214, 24
	
	\bibitem[{{Solomon} {et~al.}(2003){Solomon}, {Vanden Bout}, {Carilli}, \&
		{Guelin}}]{solomon2003}
	{Solomon}, P., {Vanden Bout}, P., {Carilli}, C., \& {Guelin}, M. 2003, \nat,
	426, 636
	
	\bibitem[{{Solomon} {et~al.}(1997){Solomon}, {Downes}, {Radford}, \&
		{Barrett}}]{solomon1997}
	{Solomon}, P.~M., {Downes}, D., {Radford}, S.~J.~E., \& {Barrett}, J.~W. 1997,
	\apj, 478, 144
	
	\bibitem[{{Solomon} \& {Vanden Bout}(2005)}]{solomon2005}
	{Solomon}, P.~M., \& {Vanden Bout}, P.~A. 2005, \araa, 43, 677
	
	\bibitem[{{Speagle} {et~al.}(2014){Speagle}, {Steinhardt}, {Capak}, \&
		{Silverman}}]{speagle2014}
	{Speagle}, J.~S., {Steinhardt}, C.~L., {Capak}, P.~L., \& {Silverman}, J.~D.
	2014, \apjs, 214, 15
	
	\bibitem[{{Spergel} {et~al.}(2007){Spergel}, {Bean}, {Dor{\'e}}, {Nolta},
		{Bennett}, {Dunkley}, {Hinshaw}, {Jarosik}, {Komatsu}, {Page}, {Peiris},
		{Verde}, {Halpern}, {Hill}, {Kogut}, {Limon}, {Meyer}, {Odegard}, {Tucker},
		{Weiland}, {Wollack}, \& {Wright}}]{spergel2007}
	{Spergel}, D.~N., {Bean}, R., {Dor{\'e}}, O., {et~al.} 2007, \apjs, 170, 377
	
	\bibitem[{{Swinbank} {et~al.}(2014){Swinbank}, {Simpson}, {Smail}, {Harrison},
		{Hodge}, {Karim}, {Walter}, {Alexander}, {Brandt}, {de Breuck}, {da Cunha},
		{Chapman}, {Coppin}, {Danielson}, {Dannerbauer}, {Decarli}, {Greve},
		{Ivison}, {Knudsen}, {Lagos}, {Schinnerer}, {Thomson}, {Wardlow}, {Wei{\ss}},
		\& {van der Werf}}]{swinbank2014}
	{Swinbank}, A.~M., {Simpson}, J.~M., {Smail}, I., {et~al.} 2014, \mnras, 438,
	1267
	
	\bibitem[{{Symeonidis} {et~al.}(2011){Symeonidis}, {Georgakakis}, {Seymour},
		{Auld}, {Bock}, {Brisbin}, {Buat}, {Burgarella}, {Chanial}, {Clements},
		{Cooray}, {Eales}, {Farrah}, {Franceschini}, {Glenn}, {Griffin},
		{Hatziminaoglou}, {Ibar}, {Ivison}, {Mortier}, {Oliver}, {Page},
		{Papageorgiou}, {Pearson}, {P{\'e}rez-Fournon}, {Pohlen}, {Rawlings},
		{Raymond}, {Rodighiero}, {Roseboom}, {Rowan-Robinson}, {Scott}, {Smith},
		{Tugwell}, {Vaccari}, {Vieira}, {Vigroux}, {Wang}, \& {Wright}}]{syme2011}
	{Symeonidis}, M., {Georgakakis}, A., {Seymour}, N., {et~al.} 2011, \mnras, 417,
	2239
	
	\bibitem[{{Symeonidis} {et~al.}(2014){Symeonidis}, {Georgakakis}, {Page},
		{Bock}, {Bonzini}, {Buat}, {Farrah}, {Franceschini}, {Ibar}, {Lutz},
		{Magnelli}, {Magdis}, {Oliver}, {Pannella}, {Paolillo}, {Rosario},
		{Roseboom}, {Vaccari}, \& {Villforth}}]{syme2014}
	{Symeonidis}, M., {Georgakakis}, A., {Page}, M.~J., {et~al.} 2014, \mnras, 443,
	3728
	
	\bibitem[{{Tacconi} {et~al.}(2010){Tacconi}, {Genzel}, {Neri}, {Cox}, {Cooper},
		{Shapiro}, {Bolatto}, {Bouch{\'e}}, {Bournaud}, {Burkert}, {Combes},
		{Comerford}, {Davis}, {Schreiber}, {Garcia-Burillo}, {Gracia-Carpio}, {Lutz},
		{Naab}, {Omont}, {Shapley}, {Sternberg}, \& {Weiner}}]{tacconi2010}
	{Tacconi}, L.~J., {Genzel}, R., {Neri}, R., {et~al.} 2010, \nat, 463, 781
	
	\bibitem[{{Tacconi} {et~al.}(2013){Tacconi}, {Neri}, {Genzel}, {Combes},
		{Bolatto}, {Cooper}, {Wuyts}, {Bournaud}, {Burkert}, {Comerford}, {Cox},
		{Davis}, {F{\"o}rster Schreiber}, {Garc{\'{\i}}a-Burillo}, {Gracia-Carpio},
		{Lutz}, {Naab}, {Newman}, {Omont}, {Saintonge}, {Shapiro Griffin}, {Shapley},
		{Sternberg}, \& {Weiner}}]{tacconi2013}
	{Tacconi}, L.~J., {Neri}, R., {Genzel}, R., {et~al.} 2013, \apj, 768, 74
	
	\bibitem[{{Thomson} {et~al.}(2012){Thomson}, {Ivison}, {Smail}, {Swinbank},
		{Weiss}, {Kneib}, {Papadopoulos}, {Baker}, {Sharon}, \& {van
			Moorsel}}]{thomson2012}
	{Thomson}, A.~P., {Ivison}, R.~J., {Smail}, I., {et~al.} 2012, \mnras, 425,
	2203
	
	\bibitem[{{Turner} {et~al.}(1990){Turner}, {Martin}, \& {Ho}}]{turner1990}
	{Turner}, J.~L., {Martin}, R.~N., \& {Ho}, P.~T.~P. 1990, \apj, 351, 418
	
	\bibitem[{{Usero} {et~al.}(2015){Usero}, {Leroy}, {Walter}, {Schruba},
		{Garc{\'{\i}}a-Burillo}, {Sandstrom}, {Bigiel}, {Brinks}, {Kramer},
		{Rosolowsky}, {Schuster}, \& {de Blok}}]{usero2015}
	{Usero}, A., {Leroy}, A.~K., {Walter}, F., {et~al.} 2015, \aj, 150, 115
	
	\bibitem[{{Vanden Bout} {et~al.}(2004{\natexlab{a}}){Vanden Bout}, {Solomon},
		\& {Maddalena}}]{vanden2004}
	{Vanden Bout}, P.~A., {Solomon}, P.~M., \& {Maddalena}, R.~J.
	2004{\natexlab{a}}, \apjl, 614, L97
	
	\bibitem[{{Wagg} {et~al.}(2005){Wagg}, {Wilner}, {Neri}, {Downes}, \&
		{Wiklind}}]{wagg2005}
	{Wagg}, J., {Wilner}, D.~J., {Neri}, R., {Downes}, D., \& {Wiklind}, T. 2005,
	\apjl, 634, L13
	
	\bibitem[{{Wei{\ss}} {et~al.}(2007){Wei{\ss}}, {Downes}, {Neri}, {Walter},
		{Henkel}, {Wilner}, {Wagg}, \& {Wiklind}}]{weiss2007}
	{Wei{\ss}}, A., {Downes}, D., {Neri}, R., {et~al.} 2007, \aap, 467, 955
	
	\bibitem[{{Wei{\ss}} {et~al.}(2003){Wei{\ss}}, {Henkel}, {Downes}, \&
		{Walter}}]{weiss2003}
	{Wei{\ss}}, A., {Henkel}, C., {Downes}, D., \& {Walter}, F. 2003, \aap, 409,
	L41
	
	\bibitem[{{Whitaker} {et~al.}(2012){Whitaker}, {van Dokkum}, {Brammer}, \&
		{Franx}}]{whitaker2012}
	{Whitaker}, K.~E., {van Dokkum}, P.~G., {Brammer}, G., \& {Franx}, M. 2012,
	\apjl, 754, L29
	
	\bibitem[{{Whitaker} {et~al.}(2011){Whitaker}, {Labb{\'e}}, {van Dokkum},
		{Brammer}, {Kriek}, {Marchesini}, {Quadri}, {Franx}, {Muzzin}, {Williams},
		{Bezanson}, {Illingworth}, {Lee}, {Lundgren}, {Nelson}, {Rudnick}, {Tal}, \&
		{Wake}}]{whitaker2011}
	{Whitaker}, K.~E., {Labb{\'e}}, I., {van Dokkum}, P.~G., {et~al.} 2011, \apj,
	735, 86
	
	\bibitem[{{Wu} {et~al.}(2005){Wu}, {Evans}, {Gao}, {Solomon}, {Shirley}, \&
		{Vanden Bout}}]{wu2005}
	{Wu}, J., {Evans}, II, N.~J., {Gao}, Y., {et~al.} 2005, \apjl, 635, L173
	
	\bibitem[{{Wu} {et~al.}(2010){Wu}, {Evans}, {Shirley}, \& {Knez}}]{wu2010}
	{Wu}, J., {Evans}, II, N.~J., {Shirley}, Y.~L., \& {Knez}, C. 2010, \apjs, 188,
	313
	
	\bibitem[{{Wuyts} {et~al.}(2011){Wuyts}, {F{\"o}rster Schreiber}, {Lutz},
		{Nordon}, {Berta}, {Altieri}, {Andreani}, {Aussel}, {Bongiovanni}, {Cepa},
		{Cimatti}, {Daddi}, {Elbaz}, {Genzel}, {Koekemoer}, {Magnelli}, {Maiolino},
		{McGrath}, {P{\'e}rez Garc{\'{\i}}a}, {Poglitsch}, {Popesso}, {Pozzi},
		{Sanchez-Portal}, {Sturm}, {Tacconi}, \& {Valtchanov}}]{wuyts2011a}
	{Wuyts}, S., {F{\"o}rster Schreiber}, N.~M., {Lutz}, D., {et~al.} 2011, \apj,
	738, 106
	
	\bibitem[{{Zavala} {et~al.}(2015){Zavala}, {Yun}, {Aretxaga}, {Hughes},
		{Wilson}, {Geach}, {Egami}, {Gurwell}, {Wilner}, {Smail}, {Blain}, {Chapman},
		{Coppin}, {Dessauges-Zavadsky}, {Edge}, {Monta{\~n}a}, {Nakajima}, {Rawle},
		{S{\'a}nchez-Arg{\"u}elles}, {Swinbank}, {Webb}, \& {Zeballos}}]{zavala2015}
	{Zavala}, J.~A., {Yun}, M.~S., {Aretxaga}, I., {et~al.} 2015, \mnras, 452, 1140
	
\end{thebibliography}
\end{document}